\documentclass[twocolumn]{aastex631}

\usepackage{graphicx}% Include figure files
\usepackage{dcolumn}% Align table columns on decimal point
\usepackage{bm}% bold math

\usepackage{color, soul}
\usepackage{amsmath}
\usepackage{braket}
\usepackage{xcolor}

\usepackage{changepage}
\newcommand{\newfootnote}[1]{\footnote{#1}}

\usepackage{hyperref}

\usepackage{booktabs, tabularx, makecell, multirow, xspace}
\renewcommand{\arraystretch}{2}
\newcolumntype{C}[1]{>{\centering\let\newline\\\arraybackslash\hspace{0pt}}m{#1}}

\definecolor{nicered}{rgb}{0.7,0.1,0.1}
\definecolor{darkorchid}{rgb}{0.6, 0.2, 0.8}

\newcommand{\dBIC}{$\Delta$BIC\,}

\usepackage[marginal]{footmisc}
\begin{document}

\title{A Critical Assessment of 
Solutions to the Galaxy Diversity Problem}

\correspondingauthor{Aidan Zentner}
\author{Aidan Zentner}
 \affiliation{Department of Physics, Princeton University, Princeton, NJ 08544, USA}
 \email{azentner@princeton.edu}

\author{Siddharth Dandavate}
 \affiliation{Department of Physics, Princeton University, Princeton, NJ 08544, USA}
 
 \author[0000-0002-5999-106X]{Oren Slone}
 \affiliation{Department of Physics, Princeton University, Princeton, NJ 08544, USA}
  \affiliation{Center for Cosmology and Particle Physics, Department of Physics,\\ New York University, New York, NY 10003, USA}

\author[0000-0002-8495-8659]{Mariangela Lisanti}
 \affiliation{Department of Physics, Princeton University, Princeton, NJ 08544, USA}
 \affiliation{Center for Computational Astrophysics, Flatiron Institute, 162 Fifth Ave, New York, NY 10010, USA}

\date{\today}

\begin{abstract}
Galactic rotation curves exhibit a diverse range of inner slopes.  Observational data indicates that explaining this diversity may require a mechanism that correlates a galaxy's surface brightness with the central-most region of its dark matter halo.  In this work, we compare several concrete models that capture the relevant physics required to explain the galaxy diversity problem.  We focus specifically on a Self-Interacting Dark Matter~(SIDM) model with an isothermal core and two Cold Dark Matter~(CDM) models with/without baryonic feedback.  In contrast to the CDM case, the SIDM model can lead to the formation of an isothermal core in the halo, and is also  mostly insensitive to baryonic feedback processes, which act on longer time-scales. Using rotation curves from 90 galaxies in the \emph{Spitzer} Photometry \& Accurate Rotation Curves~(SPARC) catalog, we perform a comprehensive model comparison that addresses issues of statistical methodology from prior works.  The best-fit halo models that we recover are consistent with standard CDM concentration-mass and abundance matching relations.  We find that both the SIDM and feedback-affected CDM models are better than a CDM model with no feedback in explaining the rotation curves of low and high surface brightness galaxies in the sample.  However, when compared to each other, there is no strong statistical preference for either the SIDM or the feedback-affected CDM halo model as the source of galaxy diversity in the SPARC catalog.
\end{abstract}

\section{\label{intro}Introduction}

Galactic rotation curves have historically served as a powerful tool for understanding the distribution of dark and visible matter in the Universe.  With recent catalogs of rotation curves that span a wide range of galaxy masses, we now have the unique opportunity to test the consistency of dark matter (DM) models across a broad population of galaxies.  One important observation that must be accounted for by any DM model is the diversity of galactic rotation curves and their apparent correlation with the baryonic properties of a galaxy \citep{Flores_1994, de_Blok_2001, Kuzio_de_Naray_2010, Oh_2011, Oman_2015}. In this work, we use the \emph{Spitzer} Photometry \& Accurate Rotation Curves~(SPARC) catalog~\citep{sparcmaster} to critically assess the preference for models of collisionless Cold DM~(CDM) with and without feedback, and for self-interacting DM (SIDM).

The diversity problem is the simple observation that the rotation curves of spiral galaxies exhibit a diverse range of inner slopes. Specifically, high surface brightness galaxies typically have more steeply rising rotation curves than do their low surface brightness counterparts, even if both have similar halo masses.  This behavior cannot easily be explained if halo formation is self-similar, for example if the halo is modeled using the standard Navarro-Frenk-White~(NFW) profile~\citep{nfwOriginal}. Within the CDM framework, two galaxies with similar halo masses will have comparable total stellar mass and concentration. If no feedback mechanism is invoked, this sets the properties of the DM profile, typically modeled as an NFW profile, such as its scale radius and overall normalization. The inner slope of the halo density is insensitive to baryonic feedback and is fixed at a constant value.    In this scenario, therefore, the halo model has little flexibility to account for correlations with the surface brightness of the galaxy~\citep{Oman_2015, katz2016, Li:2018rnd, Li:2020iib}. 

One way to address galaxy diversity within the CDM framework is to use baryonic feedback to inject energy into the halo and redistribute DM to form larger cores near the halo's center \citep{10.1093/mnras/283.3.L72, 2002MNRAS.333..299G, 10.1111/j.1365-2966.2004.08424.x, 2010Natur.463..203G, 2012MNRAS.421.3464P,DiCintio:2014xia,2014ApJ...789L..17M, 2014MNRAS.437..415D,Chan:2015tna,2016MNRAS.456.3542T,2020MNRAS.497.2393L,2020MNRAS.495L..46M}. For example, repetitive star bursts have been demonstrated to effectively produce cores in the brightest dwarf galaxies~\citep{2012MNRAS.421.3464P}.  For less massive dwarfs, such feedback is not strong enough to efficiently remove a galaxy's inner-most mass.  For larger systems, the galaxies are so massive that feedback is insufficiently powerful to create coring and profiles similar to NFW are preserved. Baryonic feedback processes therefore link the properties of a galaxy's DM density profile to its baryonic content, as needed to explain the galaxy diversity, and provide improved fits to rotation curve data~\citep{katz2016, Allaert_2017,Li:2018rnd, Li:2020iib, 2020MNRAS.495...58S}.

An alternate approach to address the galaxy diversity problem is to change the properties of the DM model itself. A well-known example is that of SIDM~\citep{Spergel:1999mh}. If DM can self interact, heat transfer in the central regions of the halo is efficient, creating an isothermal region within the density profile \citep{Vogelsberger_2012, Zavala_2013, Rocha_2013, Peter_2013}. This central, isothermal profile is exponentially dependent on the baryonic potential of the galaxy. While a DM-dominated system will form a core, a baryon-dominated system will retain a more cuspy center, similar to that of an NFW profile~\citep{kap2014}. Thus, in the SIDM framework, the size of the core naturally correlates with the baryonic surface brightness of the galaxy, allowing for a mechanism that can account for galaxy diversity~\citep{Kamada_2017, 2017MNRAS.468.2283C, kap2019, kap2020}. 

When the time-scale for forming the isothermal core is shorter than the time-scale for baryonic feedback, SIDM halos are insensitive to the details of feedback processes \citep{kap2019}.  In this case, an SIDM halo thermalizes fast enough that the resulting density distribution is only sensitive to the present-day baryonic distribution, and not the formation history of the galaxy.   These results have been confirmed with hydrodynamic simulations, which---for the self-interaction cross sections studied---have found that SIDM density profiles are more robust to stellar feedback than their CDM counterparts~\citep{2014MNRAS.444.3684V,2015MNRAS.452.1468F, 2017MNRAS.472.2945R,Robertson:2017mgj, 2021MNRAS.507..720S,2021arXiv210414069V,2022MNRAS.tmp..969B}.

This paper presents a systematic comparison of three models with equal numbers of degrees-of-freedom using SPARC galaxies. These models attempt to capture the important physics of: 1) the isothermal core in SIDM, 2) baryonic feedback in  CDM, and 3) CDM with no baryonic feedback. Our analysis highlights aspects of the SPARC dataset that can potentially bias the model comparison and also addresses several issues with the statistical methodology of previous SPARC rotation curve studies~\citep{katz2016, Li:2018rnd, kap2020, Li:2020iib}. We find that the CDM model with no baryonic feedback is strongly disfavored for low surface brightness galaxies, although less-so for high surface brightness galaxies.  Additionally, we find no strong statistical preference for the SIDM model or the CDM model with baryonic feedback, for either the low or high surface brightness systems.

The paper is organized as follows. In Sec.~\ref{SPARC}, we discuss the SPARC dataset, emphasizing several aspects of the publically available data that can systematically affect the results of this study.  Sec.~\ref{methods} describes the data selection, theory modeling, and likelihood analysis. Sec.~\ref{results:individual} presents the rotation curve fits to a few example galaxies and discusses over-fitting issues to the data. Sec.~\ref{results:models} presents the statistical model comparison for the scenarios considered in this work. We then conclude in Sec.~\ref{conclusions}.  App.~\ref{app:corner} provides rotation curve fits and corner plots for example galaxies spotlighted in the text. App.~\ref{app:B} focuses on several anomalous galaxies in the SPARC dataset.

Rotation curve fits and corner plots for all galaxies used in this study are publicly available on \texttt{github}.\newfootnote{\href{https://github.com/siddanda/DarkModels}{\texttt{https://github.com/siddanda/DarkModels}}} 

\section{\label{SPARC}The SPARC Dataset}

The SPARC database includes 3.6~$\mu$m surface photometry measurements and H1/H$\alpha$ rotation curves for 175 late-type galaxies~\citep{sparcmaster}. This comprehensive sample includes galaxies with stellar masses $3\times 10^7 \lesssim M_\star \lesssim 3\times 10^{11}~M_\odot$ and a wide range of luminosities, morphologies, and surface brightnesses.  While it is representative of nearby disk galaxies, it does not form a statistically complete sample~\citep{katz2016, sparcmaster}.

The SPARC catalog is publicly available and provides the observed rotation curve measurements, $V_{i,{\rm obs}}$, and their errors, $\delta V_{i,{\rm obs}}$, for each galaxy, where the subscript $i$ corresponds to the radius at which the measurement is reported. The baryonic mass models are given as the circular velocities for the disk ($V_{i,{\rm disk}}$), bulge ($V_{i,{\rm bul}}$), and gas ($V_{i,{\rm gas}}$) components. These velocities were found by solving the Poisson equation for the observed surface brightness at 3.6~$\mu$m for disk and bulge contributions and from the HI surface brightness for the gas contribution, assuming mass-to-light ratios of unity for all, $\Upsilon_{\star, \rm{disk}}=\Upsilon_{\star, \rm{bul}}=1 \, M_\odot/L_\odot$~\citep{sparcmaster}.

If a model predicts a galaxy's rotation curve contribution from DM at a certain radius, $V_{\rm DM}$, these can be used to infer the galaxy's predicted total rotation curve through,
\begin{eqnarray}
 V_{\rm tot}^2  = && V_{\rm DM}\,|V_{\rm DM}| + V_{\rm gas}\, |V_{\rm gas}| +\Upsilon_{\star, \rm{disk}} V_{\rm disk} \, |V_{\rm disk}| \nonumber \\ 
 && \hspace{0.1in}+\, \Upsilon_{\star, \rm{bul}} V_{\rm bul} \, |V_{\rm bul}| \, ,
 \label{eq:totv}
\end{eqnarray}
where the $i$ subscripts have been suppressed for notational simplicity.\newfootnote{Absolute values are needed to account for negative velocities that are provided in the SPARC catalog---see~\cite{sparcmaster}.}  The model can then be compared to data to quantify its goodness of fit, as was done in \cite{katz2016, Kamada_2017, Li:2018rnd, kap2019, kap2020, Li:2020iib}.

There are three important aspects of the SPARC data that should be carefully considered when performing such statistical model comparisons.  First, uncertainties on the baryonic contribution to rotation curves are not included in the SPARC catalog.  Technically, such uncertainties should be directly propagated to the predicted $V_{\rm tot}$ through $V_{\rm gas}$, $V_{\rm disk}$ and $V_{\rm bul}$ in Eq.~\eqref{eq:totv}.  
In line with other studies performed to date, we are unable to properly account for these uncertainties given the information that is publicly available in the SPARC catalog. Moreover, in the case of the SIDM and feedback-affected CDM model considered here, the predicted value of $V_{\rm DM}$ itself depends on the galaxy's baryonic distribution and the uncertainties in baryonic velocities should in principle be propagated into the DM model.  Unfortunately, it is not possible to estimate how this affects the conclusions of this study  without knowing the magnitude of the baryonic uncertainties.

Second, some models may over-fit the data.  Many SPARC galaxies have data points whose spread around the best-fit model predictions is smaller than expected from the uncertainties on those data points.  This could be indicative of correlations between data points that are not accounted for in the model, or of over-estimation of errors for those data points.  To address this issue, potentially problematic galaxies will be flagged when presenting results. The selection tests for over-fit galaxies are discussed in detail in Sec.~\ref{results:individual}. 

Lastly, due to the possibility of unknown systematic biases in the selection of SPARC galaxies, one might be concerned that the best-fit parameters of a Bayesian analysis will be strongly prior-dependent.  For this reason, the benchmark analysis of this study is designed with reasonably loose priors.  Sec.~\ref{results:models} includes a summary of how variations on these priors affect the primary conclusions.

\section{\label{methods}Methodology}

From the SPARC catalog, 98 galaxies with at least 10 data points are selected that have a well-defined velocity for the flat part of the rotation curve~\citep{2016ApJ...816L..14L}, $V_\text{f}$, with inclination $i \geq 30^\circ$, as well as those suitable for dynamical studies with quality flag $Q\neq3$. For each galaxy, our analysis uses the observed circular velocity and associated uncertainty as a function of galactocentric radius, as well as the derived contributions to the total velocity from the gas, galactic disk, and galactic bulge, if present. 

The following subsections describe how we model mass distributions for CDM~(Sec.~\ref{methods:cdm}) and SIDM~(Sec.~\ref{methods:sidm}) halos and how the predicted velocity rotation curves are then fit to data~(Sec.~\ref{methods:likelihood}).  Throughout, we work with virial quantities and use the Planck 2015 cosmology with $H_0 = 67.8$~km~s$^{-1}$~Mpc$^{-1}$, $\Omega_m = 0.308$, and $\Omega_\Lambda = 0.692$~\citep{planck2015}. The virial mass, $M_{\rm vir}$, is defined as the enclosed halo mass at the virial radius, $r_{\rm vir}$, where the average enclosed density is a factor $\Delta_c$ greater than the critical density.  Following \cite{1998ApJ...495...80B} and assuming the Planck cosmology, $\Delta_c \approx 102$. 

\subsection{\label{methods:cdm} Mass Models for Cold Dark Matter}

Two spherically-symmetric halo models for CDM are considered, one with and one without baryonic feedback processes accounted for.  Both CDM halos are modeled using the $(\alpha, \beta, \gamma)$ density profile, 
\begin{equation}
    \rho_{\scriptscriptstyle{\rm CDM}}(r) = \frac{\rho_s}{\left(r/r_s\right)^\gamma \left[1 + \left(r/r_s\right)^\alpha\right]^{(\beta-\gamma)/\alpha}} \, ,
    \label{eq:cdm}
\end{equation}
where $\rho_s$ is the scale density and $r_s$ is the scale radius~\citep{1990ApJ...356..359H}.  The logarithmic slope of the density profile goes as $d \log\rho / d\log r = -2$ at a galactocentric distance of  
\begin{equation}
    r_{\scriptscriptstyle{-2}} = \left( \frac{2-\gamma}{\beta-2} \right)^{1/\alpha} r_s \, .
    \label{eq:slope}
\end{equation}
The enclosed mass, $M(r)$, takes an analytic form for this profile  and the prediction for the DM contribution to the circular velocity at any radius is 
\begin{equation}
   V_{\rm DM}^2(r) = \frac{G M(r)}{r} \, ,
   \label{eq:V_Mr}
\end{equation}
where $G$ is Newton's gravitational constant.  Once the $\alpha, \beta, \gamma$ exponents are specified, there are two free parameters remaining. Without loss of generality, they are chosen to be the concentration, $c_{\rm vir} = r_{\rm vir}/r_{\scriptscriptstyle{-2}}$, and the virial velocity,
\begin{equation}
    V_{\rm vir} = r_s H_0 c_{\rm vir} \sqrt{\frac{\Delta_c}{2}} \, .
\end{equation}
To predict the total rotational velocity $V_{\rm tot}$ and compare to data, the mass-to-light ratio parameters, $\Upsilon_{\star, \rm{disk}}$ and $\Upsilon_{\star, \rm{bul}}$, must also be specified.  Thus, the CDM models considered here have a total of four free parameters.

The baseline CDM scenario considered in this work assumes no baryonic feedback.  In this case, the density distribution is given by the NFW profile with $(\alpha, \beta, \gamma) = (1, 3, 1)$ and $r_{\scriptscriptstyle{-2}} = r_s$.  This case is also a good proxy for CDM scenarios with weak baryonic feedback.

As an example of a feedback-affected CDM halo, we use the DC14 model~\citep{DiCintio:2014xia,2014MNRAS.437..415D}, which is based on galaxies of dwarf to Milky Way masses from the Making Galaxies In a Cosmological Context~(MAGICC) project~\citep{10.1093/mnras/sts028}. These hydrodynamic MAGICC simulations account for stellar feedback, which can lead to a cored density profile depending on the galaxy's stellar-to-halo mass, $M_\star/M_{\rm vir}$.  In this case, the $(\alpha, \beta, \gamma)$ exponents are:
\begin{eqnarray}
\alpha &=& 2.94 - \log_{10}\left[\left(10^{X+2.33}\right)^{-1.08} + \left( 10^{X+2.33} \right)^{2.29}\right]\nonumber  \\ 
\beta &=& 4.23 + 1.34 X + 0.26 X^2 \nonumber \\
\gamma &=& -0.06 + \log_{10}\left[ \left(10^{X+2.56}\right)^{-0.68} + 10^{X+2.56}\right] \, , \nonumber
\end{eqnarray}
with $X = \log_{10}\left(M_\star/M_\text{vir}\right)$.  The DC14 model is valid in the mass range $-4.1 < \log_{10}(M_\star/M_{\rm vir}) < -1.3$.  Because the model does not capture potential feedback from active galactic nuclei~(AGN) at masses above this range, these galaxies are excluded from our analysis after the fitting procedure is completed, further reducing the sample to 90 galaxies from 98.  $M_\star$ is estimated by using a combination of the total and bulge luminosities reported in the SPARC catalog, $L_{\rm tot}$ and $L_{\rm bul}$, and the mass-to-light ratio parameters,
\begin{equation}
    M_\star \approx \Upsilon_{\star, \rm{disk}} \left( L_{\rm tot} - L_{\rm bul} \right) + \Upsilon_{\star, \rm{bul}} L_{\rm bul} \, ,
\end{equation}
under the assumption that the gas is a negligible contribution to $L_{\rm tot}$.

The overall properties of the DC14 model have been confirmed with other hydrodynamic simulation codes that include stellar feedback~\citep{Chan:2015tna,2016MNRAS.456.3542T,2020MNRAS.497.2393L,2020MNRAS.495L..46M}.  In general, core-formation is suppressed for $M_\star/M_{\rm vir} \lesssim 10^{-4}$ and becomes maximal when approaching $M_\star/M_{\rm vir} \sim 5\times10^{-3}$.  For $M_\star/M_{\rm vir} \gtrsim 10^{-2}$, the halo density profile becomes preferentially cuspy once again, though the degree to which this occurs depends on AGN  feedback~\citep{2020MNRAS.495L..46M}.  The details on how rapidly the transitions between cuspy and cored occur as a function of $M_\star/M_{\rm vir}$, as well as the expected degree of scatter about the median distribution, does vary between simulation codes.  

\cite{Bose:2018oaj,  Dutton:2018nop, 2019MNRAS.488.2387B, Dutton:2020vne} emphasize that bursty star formation histories are not the only requirement for core formation.  Additionally, one needs a star formation threshold that is large enough to allow enough gas to build up near the center of the galaxy, so that when it is blown out by supernova explosions, it can effectively perturb the DM orbits away from the center of the galaxy.  As a result, simulation codes with lower star formation thresholds do  not see the same level of coring in dwarf systems as captured by the DC14 model.

\subsection{\label{methods:sidm} Mass Model for SIDM}

To model the SIDM density distribution, we follow the analytic approach of \cite{Rocha:2012jg,Kaplinghat:2013xca,kap2016}, which has been verified with numerical simulations of isolated SIDM halos~\citep{2017MNRAS.468.2283C, 2018ApJ...853..109E, 2018MNRAS.479..359S}.  This procedure requires dividing the halo into an inner and an outer region, separated at a galactocentric radius $r_1$ where DM particles average one interaction over the age of the galaxy. Beyond $r_1$, SIDM interactions are negligible, and the DM profile is expected to resemble the CDM result, modeled using the NFW profile. Below $r_1$, the self interactions thermalize the DM distribution, and the density distribution is well-modeled as an isothermal profile,  
\begin{equation}
    \rho_{\rm iso}(r) = \rho_{\text{0}} \exp(-\Delta\Phi_{\text{tot}}(r)/\sigma_{\text{v0}}^2) \, ,
    \label{eq:sidm1}
\end{equation}
where $\Delta \Phi_{\text{tot}} \equiv \Phi_{\rm tot}(r) - \Phi_{\rm tot}(0)$ is the total gravitational potential, $\rho_{0}$ is the central DM density, and $\sigma_\text{v0}$ is the one-dimensional velocity dispersion.  Thus, the SIDM density profile is given by
\begin{equation}
\rho_{\rm SIDM}(r) = 
  \begin{cases}
    \rho_{\rm iso}(r) & r < r_1 \\
    \rho_{\scriptscriptstyle{\rm NFW}}(r) & r \geq r_1 \, .
  \end{cases}
  \label{eq:sidm2}
\end{equation}
The NFW and isothermal distributions are ``stitched'' together at $r_1$ such that the enclosed mass and density are continuous.  Additionally, the value of $r_1$ is set by the relation
\begin{equation}
    \rho_{\rm iso} (r_1) \, t_{\text{age}} \braket{\sigma \, v_{\text{rel}}} / m_\chi = 1 \, ,
    \label{eq:sidm3}
\end{equation}
 where $\sigma/m_\chi$ is the self-scattering cross section per DM particle mass, $v_{\text{rel}}$ is the relative velocity of the scattering particles, and $t_{\text{age}}$ is the age of the galaxy.  For an isothermal density distribution and Maxwellian velocity distribution, $\braket{\sigma v_\text{rel}} = \sigma (4 / \sqrt{\pi}) \sigma_\text{v0}$.

\begin{table*}[t]
\footnotesize
\begin{center}
\begin{tabular}{C{2.5cm} | C{3cm}C{2cm}C{2.5cm}C{6cm}}
  \Xhline{3\arrayrulewidth}
\renewcommand{\arraystretch}{1.2}
   & \textbf{Parameter} & \textbf{Units} & \textbf{Prior Type}     &  \textbf{Prior Range}  \\  
   \hline
 NFW, DC14        & $V_{\rm vir}$        & km/s         & uniform   &  [0, 500]  \\
SIDM & $\Gamma_0$       & $0.1~\text{Gyr}^{-1}$        & log-uniform   &  [2, 10$^5$]  \\
     & $\sigma_\text{v0}$     & km/s              &      uniform &  [2, 500] \\ 
All Models & $\log_{10}c_{\rm vir}$                             & -- & normal    & Eqs.~\eqref{eq:cvir},~\eqref{eq:cvirdc14}\\     
 & $\Upsilon_{\star,\text{disk}}$     & $M_\odot/L_\odot$                & log-normal   &  $\mu = 0.5, \sigma = 0.1~\text{dex}$  \\
     & $\Upsilon_{\star, \text{bul}}$       & $M_\odot/L_\odot$                & log-normal   &  $\mu = 0.7, \sigma = 0.1$~\text{dex}  \\
     &  $V_\text{max}/V_\text{f}$      &  --   & uniform &  [2$^{-0.5}$, 2$^{0.5}$] \\ 
     & $M_{\rm b, tot}/M_{\rm vir}$ & --     & uniform & $[0,\ 0.2]$ \\
 \hline
  \Xhline{3\arrayrulewidth}
\end{tabular}
\end{center}
\caption{Priors on the free parameters used to model the dark matter and baryonic contributions to the rotation curve data from the SPARC dataset.  The free parameters for the NFW and DC14 models are the virial velocity, $V_{\rm vir}$, and concentration, $c_{\rm vir}$.  The prior on $c_{\rm vir}$ follows the best-fit concentration-mass relation of \cite{Dutton_2014}.  The free parameters for the SIDM model are $\Gamma_0$ and  $\sigma_\text{v0}$, as defined in Eqs.~\eqref{eq:sidm1} and \eqref{eq:sidm4}; the prior on concentration is also enforced for SIDM halos.  In all cases, we place a prior motivated by~\cite{Li:2020iib}, and references therein, on the mass-to-light ratios of the stellar disk and bulge, $\Upsilon_{\star,\text{disk}}$ and $\Upsilon_{\star,\text{bul}}$, as well as a regularization prior on $V_{\rm max}/V_\text{f}$.  The benchmark analysis also requires that the ratio of total baryonic to DM mass, $M_{\rm b, tot}/M_{\rm vir}$, be less than $0.2$.  Sec.~\ref{results:models} discusses how the conclusions of the study are affected by variations on these benchmark priors. }
\label{tab:priors}
\end{table*}

As a concrete example, we set $t_{\text{age}} = 10\ \text{Gyr}$ and $\sigma / m_\chi = 3\ \text{cm}^2/\text{g}$, but note that the resulting profile is only weakly dependent on these choices~\citep{Elbert:2014bma, Sokolenko:2018noz, kap2019}. This can be understood intuitively because typically $r_1 \approx r_{-2}$ of the corresponding NFW profile and so there is an approximate scaling $r_1 \propto \sqrt{t_{\rm age} \times \sigma/m_\chi}$. Therefore, $r_1$ is only weakly dependent on $t_{\rm age} \times \sigma/m_\chi$.  Because $r_1$ depends only on the product of $t_{\rm age}$ and $\sigma/m_\chi$, there is an inherent degeneracy between the two values. 
 
Given a baryonic density profile, $\rho_\text{b}(r)$, the isothermal SIDM profile can be found via Eqs.~(\ref{eq:sidm1}--\ref{eq:sidm3}) together with the Poisson equation,
\begin{equation}
    \nabla^2 \Phi_{\text{tot}}(r) = 4\pi G \left[\rho_{\rm iso}(r) + \rho_\text{b}(r)\right] \, .
    \label{eq:sidm2}
\end{equation}
To obtain an estimate for $\rho_b(r)$, the baryonic mass profile is taken to be spherically symmetric with total mass at any radius $r_i$ (where a measurement is reported) given by, 
\begin{eqnarray}
    M_\text{b}(r_i) = &&(\Upsilon_{\star, \text{disk}}V_\text{disk}\,|V_\text{disk}| + \Upsilon_{\star, \text{bul}} V_\text{bul}\,|V_\text{bul}| \nonumber \\
    && \hspace{0.1in} + \, V_\text{gas}\,|V_\text{gas}|)r_i/G \, .
    \label{eq:M_b}
\end{eqnarray}
An approximate density profile is then created as follows. A function for $M_{\rm b}(r)$ is created by quadratic interpolation between the smallest and largest radii, $r_{\rm min}$ and $r_{\rm max}$, at which measurements exist. The density profile is then taken to be,
\begin{equation}
\rho_{\text{b}}(r) = 
  \begin{cases}
    M_{\text{b}}(r_{\rm min})/(4\pi r_{\rm min}^3 / 3) & r < r_{\rm min} \\
    \left(d M_{\rm b}(r)/dr\right) / \left(4 \pi r^2\right) & r_{\rm min} \leq r \leq r_{\rm max} \, \\
    \rho_{\text{b}}(r_{\rm max}) \left(r/r_{\rm max}\right)^{-4} & r_{\rm max} < r \, .
  \end{cases}
  \label{eq:rho_b}
\end{equation}
The rotational velocity associated with the SIDM halo follows from $\rho_{\rm SIDM}$ through Eq.~\eqref{eq:V_Mr}.  

There are four free parameters for the SIDM model, which are chosen to be $\Gamma_0$, $\sigma_\text{v0}$, $\Upsilon_{\star, \rm{disk}}$ and $\Upsilon_{\star, \rm{bul}}$.  As in \cite{kap2019}, the scattering rate is defined as, 
\begin{equation}
\Gamma_0 = \rho_0 (\sigma / m_\chi) (4 / \sqrt{\pi}) \sigma_\text{v0}. 
\label{eq:sidm4}
\end{equation}
We do not consider a feedback-affected SIDM halo, having confirmed that the best-fit thermalization timescale is typically less than 0.25~Gyr for most of the SPARC galaxies in our sample. Feedback could potentially change our results if it acts on a timescale which is much shorter than this. However, as mentioned above, simulations show that this does not typically occur, even for larger cross sections than $3$~cm$^2$/g.

\subsection{\label{methods:likelihood} Likelihood Procedure}

For a model with free parameters $\boldsymbol{\theta}$, the log-likelihood, $\ln{\mathcal{L}}(\boldsymbol{\theta})$, is defined to be,
\begin{equation}
    \ln{\mathcal{L}}\left(\boldsymbol{\theta}\right) = -\frac{1}{2} \sum_i \left(\frac{V_{i, \text{obs}} - V_{ \text{tot}}(\boldsymbol{\theta},r_i)}{\delta V_{i,\text{obs}}} \right)^2 \, ,
\end{equation}
where $V_{i, \text{obs}}$ and $\delta V_{i, \text{obs}}$ are the velocity and associated uncertainty of the $i^\text{th}$ data point in a galaxy's rotation curve, and $V_{\text{tot}}(\boldsymbol{\theta},r_i)$ is the corresponding model prediction at radius $r_i$. The parameter space is scanned using the affine-invariant Markov Chain Monte Carlo~(MCMC) sampler, \texttt{emcee}~\citep{emceeBase, emcee}.  We use 48 walkers and 20,000 steps, with a burn-in period of 5,000 steps.  The convergence of the post burn-in chain for each parameter is verified by ensuring that its length is at least 50 times the estimated  autocorrelation time~\citep{emceeBase, emcee}. A limited number of galaxies fail this test, oftentimes if a parameter's posterior distribution is not unimodal. These galaxies are run for an additional 10,000 steps.  Those that are still not converged after this step are NGC6015 for NFW; none for DC14; and NGC0247 for SIDM. 

We perform this sampling in total for three different models: NFW, DC14, and SIDM.  The priors used for the benchmark analysis in this study are summarized in Tab.~\ref{tab:priors}.  All the models considered share the same prior on the disk and bulge mass-to-light ratios and have total baryonic mass to DM mass $M_{\rm b, tot}/M_{\rm vir} \leq 0.2$.  Additionally, as in \cite{kap2019}, we restrict $2^{-0.5} \leq V_{\rm max}/V_{\rm f} \leq 2^{0.5}$ to enforce that the predicted flat rotation curve corresponds to the measured $V_{\rm f}$. 

\begin{figure*}[t] %  figure placement: here, top, bottom, or page
    \centering
    \includegraphics[width=0.9\textwidth]{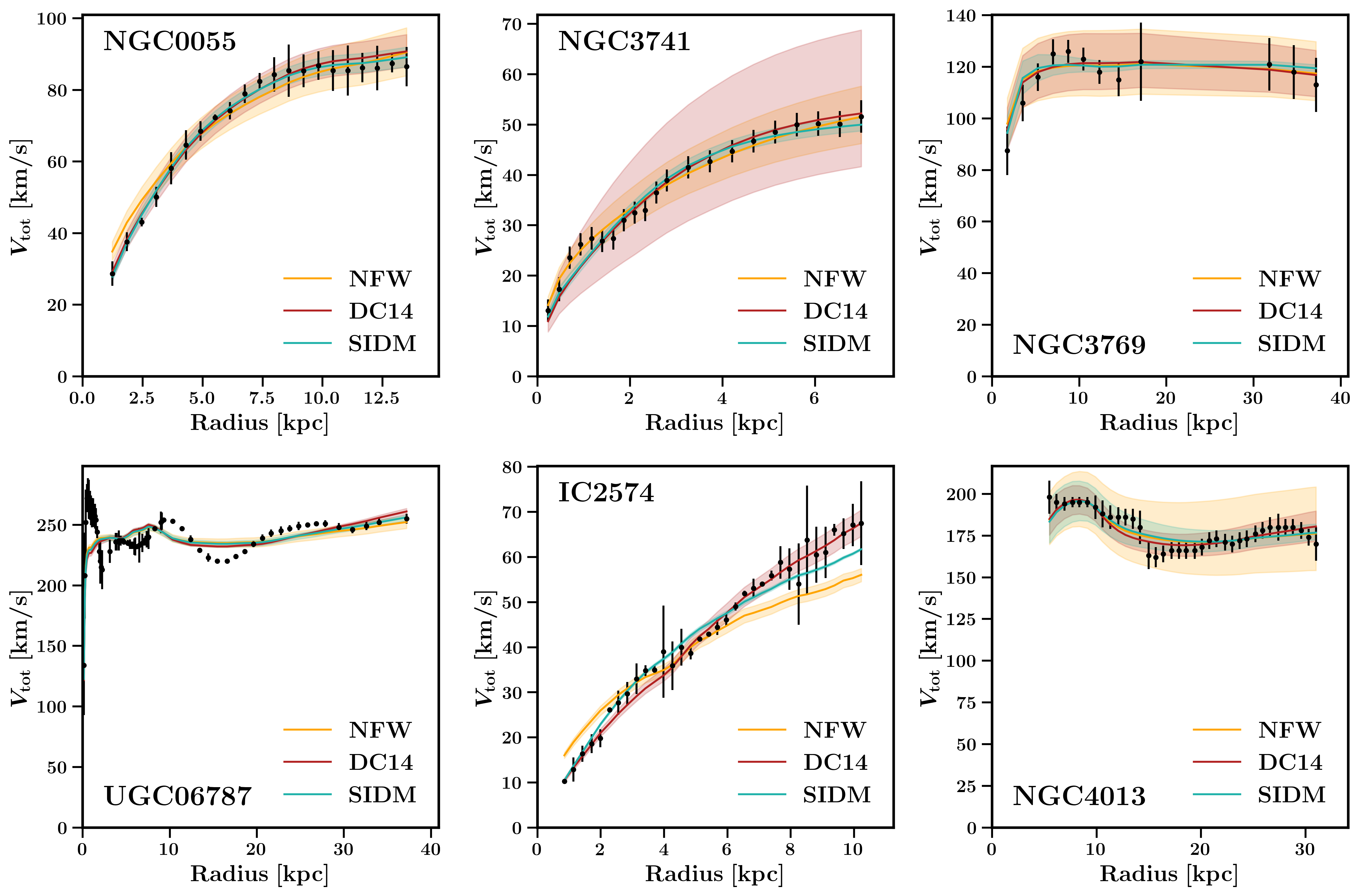}
   \caption{Rotation curve data for six example galaxies in the SPARC catalog. Curves for the NFW, DC14, and SIDM models are indicated by the solid yellow, red, and aqua lines, respectively.  In each case, the central curve and colored band correspond to the 16--50--84$^{\rm th}$ percentile of the posterior distribution.  Note the change in axis range between panels.  \emph{(Top Row)} NGC0055 is a low surface brightness galaxy in the SPARC sample where the SIDM and DC14 model do equally well and are both better fits than the NFW model.  All three models fit the data well for NGC3741 and NGC3769, which are examples of low and high surface brightness galaxies, respectively.  
   \emph{(Bottom Row)} When performing a model comparison using rotation curve data, there can be considerable galaxy-to-galaxy scatter.  It is important to keep in mind that the variability of the data between galaxies can contribute to this and to therefore look for trends across an ensemble of galaxies.  Provided here are three illustrative examples: the UGC06787 curve is highly non-monotonic, the IC2574 curve does not include much data in the region where it is expected to flatten, and the NGC4013 curve does not extend down to low radii. The relevant rotation curve fits and corner plots for these six galaxies are provided in App.~\ref{app:corner}; other relevant summary plots can be found for the remaining SPARC galaxies on the \texttt{github}.}
   \label{fig:rotationcurve}
\end{figure*}

In addition, for all models a prior is placed on the halo concentration-mass relation, assuming that it follows \cite{Dutton_2014} with 
\begin{eqnarray}
\log_{10}c_{\rm vir} = 1.025 - 0.097 \log_{10}\left( M_{\rm vir}/\left[10^{12} h^{-1} M_\odot\right]\right) \, , \nonumber \\
\label{eq:cvir}
\end{eqnarray}
and intrinsic scatter $0.11$~dex. The concentration of a DC14 halo, $c_{\rm vir}^{\scriptscriptstyle{\rm DC14}}$, is defined in terms of the NFW concentration using
\begin{equation}
c_{\rm vir}^{\scriptscriptstyle{\rm DC14}} =  c_{\rm vir} \, \left( 1 + 0.00001 e^{3.4 \left( X + 4.5 \right)} \right) \, ,
\label{eq:cvirdc14}
\end{equation}
as provided in~\cite{DiCintio:2014xia}, but since updated.\newfootnote{Private communication with A.~DiCintio.}

Sec.~\ref{results:models} explores several variations on these benchmark priors and their effects on the conclusions of the study.

\section{\label{results:individual}Best-Fit Rotation Curve Examples}

The procedure described in Sec.~\ref{methods} has been performed for all 90 galaxies within the SPARC catalog that pass the quality cuts. For each galaxy, and for each of the three models considered in this study, we recover posteriors of the free parameters and a best-fit rotation curve.  Fig.~\ref{fig:rotationcurve} provides examples of six galaxies from the SPARC catalog together with their best-fit rotation curves. These examples are not meant to represent average trends, but rather have been chosen to highlight some relevant aspects of our analysis.  The full set of rotation curves are publicly available (see Footnote 1).

Many galaxies in the dataset prefer the DC14 or SIDM models over NFW, while other galaxies show either no preference or some preference for NFW.  These results may depend on a galaxy's surface brightness, defined as $\Sigma_0 = \Upsilon^{-1} M_\star/(2\pi R_\text{d}^2)$, where $R_\text{d}$ is the disk scale radius and the mass-to-light ratio is taken to be $\Upsilon=1 M_\odot/L_\odot$.  Throughout, we consider low surface brightness galaxies to be those with $\log_{10}\left(\Sigma_0/\left(L_\odot/\text{pc$^2$}\right)\right) < 2.5$ and high surface brightness galaxies to be those with values above this range.  

In the top row of Fig.~\ref{fig:rotationcurve}, NGC0055 is an example of a low surface brightness galaxy where clearly the NFW model struggles to reproduce the measured rotation curve at small galactic radii. However, NGC3741 is a counter example where all three models fit the data well.  The galaxy NGC3769 is a high surface brightness galaxy where again there is no strong preference for any of the three models considered.

In general, we observe galaxy-to-galaxy variation in the recovered model preferences, which underscores the importance of studying an ensemble of galaxies.   One contributing factor to this spread is the fact that some galaxies have more/less rotation curve data available and---in some cases---unusually shaped curves.  The second row of Fig.~\ref{fig:rotationcurve} provides some illustrative examples.  UGC06787 is an example of a galaxy whose rotation curve is highly non-monotonic, while the reported uncertainties for the rotation values are remarkably small. Clearly, none of the models considered in this study are able to capture this behavior. IC2574 is an example of a galaxy that passes the cut of a well-defined $V_\text{f}$, however clearly is not measured out to the radius at which the rotation becomes constant. Finally, NGC4013 is an example of a galaxy whose inner-most measurement is only at $\sim 5$~kpc and thus does not help distinguish any model where differences become apparent closer to the galactic center.

Comparing the various models requires specifying a test statistic. Prior rotation curve studies have performed model comparisons using the reduced chi-squared test~\citep{katz2016, Li:2018rnd, kap2019, kap2020, Li:2020iib}.  In particular, the chi-squared per degree of freedom, $\chi^2_\nu$, was evaluated for each galaxy in the SPARC sample, for a model of interest.  In some cases, the cumulative distribution function (CDF) of the reduced chi-squared values was then plotted.  A model was considered to be preferred over another if it resulted in more galaxies with lower $\chi^2_\nu$. There are, however, several fundamental issues with comparing models in this fashion.

The most important of these issues is that the reduced chi-squared is not an accurate measure of the goodness-of-fit for the rotation curve models under consideration.  The problem lies in how the degrees of freedom, $\nu$, are determined~\citep{2010arXiv1012.3754A}. By naive degree-of-freedom counting, one usually defines $\nu$ to be the difference between the number of data points in the sample, $N$, and the number of fit parameters, $P$.  However, as highlighted in \cite{2010arXiv1012.3754A}, this is only correct for a linear model, $f\left(\boldsymbol{\theta},\boldsymbol{x}\right)$, defined as
\begin{equation}
    f\left(\boldsymbol{\theta},\boldsymbol{x}\right) = \sum_{p=1}^{P} \theta_p B_p\left(\boldsymbol{x}\right)  \, ,
    \label{eq:linear}
\end{equation}
where $\theta_p$ are the free parameters and $B_p\left(\mathbf{x}\right)$ are basis functions.  If the basis functions are linearly independent, then the number of degrees of freedom is given by $\nu = N-P$.  If this requirement fails, then determining $\nu$ is highly non-trivial.  Indeed, for non-linear models, $\nu \in [0, N-1]$ and may not even remain constant during the fitting procedure~\citep{2010arXiv1012.3754A}.  The NFW, DC14, and SIDM models defined in Secs.~\ref{methods:cdm} and~\ref{methods:sidm} are non-linear as they clearly do not take the form of Eq.~\eqref{eq:linear}.  While the chi-squared can still be robustly evaluated---and thus the likelihood procedure remains intact---the \emph{reduced} chi-squared is not an adequate measure of the goodness-of-fit. 

\begin{figure*}[t] %  figure placement: here, top, bottom, or page
    \centering
    \includegraphics[width=0.8\textwidth]{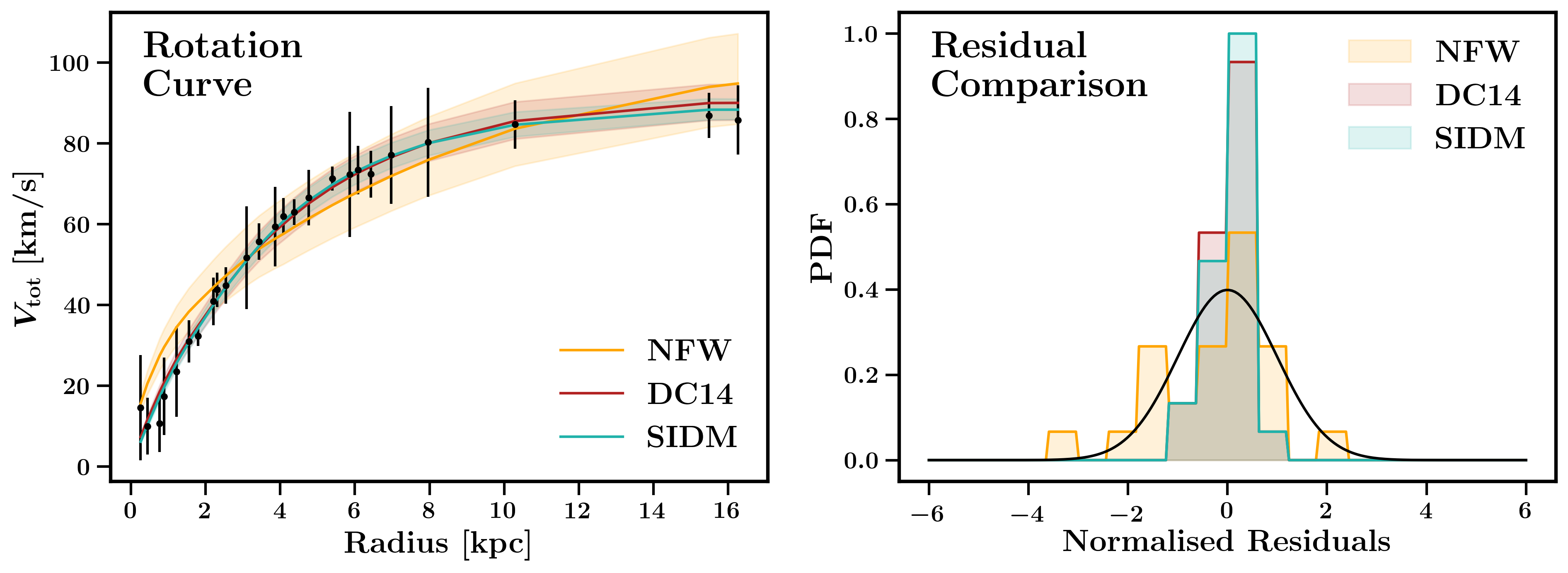}
   \caption{Rotation curve data for the SPARC galaxy F583-1 (black points).  The curves/bands for each halo mass model indicate the 16--50--84$^{\rm th}$ percentiles of the posterior distribution.  The NFW, DC14, and SIDM models are indicated by the yellow, red, and aqua lines/bands, respectively.  \emph{(Right)}  The normalized residuals for the SIDM, DC14, and NFW models.  The solid black curve corresponds to a Gaussian distribution with mean $\mu=0$ and standard deviation $\sigma^2=1$.  If the distribution of residuals has $\sigma^2 < 1$, as is the case for the SIDM and DC14 fits to this galaxy, then it indicates an issue with over-fitting the data.}
   \label{fig:overfitting}
\end{figure*}

On its own, this observation makes it impossible to properly interpret the CDF of $\chi^2_\nu$. 
We should also briefly note, however, that even if the use of reduced chi-squared were appropriate, there would be further issues in the statistical interpretations made in~\cite{katz2016, Li:2018rnd, kap2020, Li:2020iib} that rely on  $\chi^2_\nu$.  First, models with $\chi^2_\nu \ll 1$ are indicative of over-fitting and thus should not be interpreted as good fits to the data. Second, model comparisons should be evaluated on a galaxy-by-galaxy basis by considering the change in chi-squared, or $\Delta \chi^2$, between the different model fits. However, such comparisons are only valid when comparing nested models,\newfootnote{Two models are considered nested if one includes all the fit parameters of the other.} which is not the case for the NFW, SIDM, and DC14 cases.  

To interpret the goodness-of-fit of each model to the rotation curve data, we use the distribution of normalized residuals,  
\begin{equation}
    R_i(\boldsymbol{\theta}) = \frac{V_{i, \text{obs}} - V_{\rm tot}(\boldsymbol{\theta},r_i)}{\delta V_{i,\text{obs}}} \, ,
\end{equation}
evaluated for all $i$ data points.  The distribution of $R_i$ for the true model will follow a Gaussian distribution with mean $\mu = 0$ and standard deviation $\sigma^2 = 1$.  Just because a model results in residuals that follow this Gaussian distribution does not mean that it is necessarily the true model, only that it is a viable candidate.

\begin{figure*}[t] %  figure placement: here, top, bottom, or page
    \centering
    \includegraphics[width=0.9\textwidth]{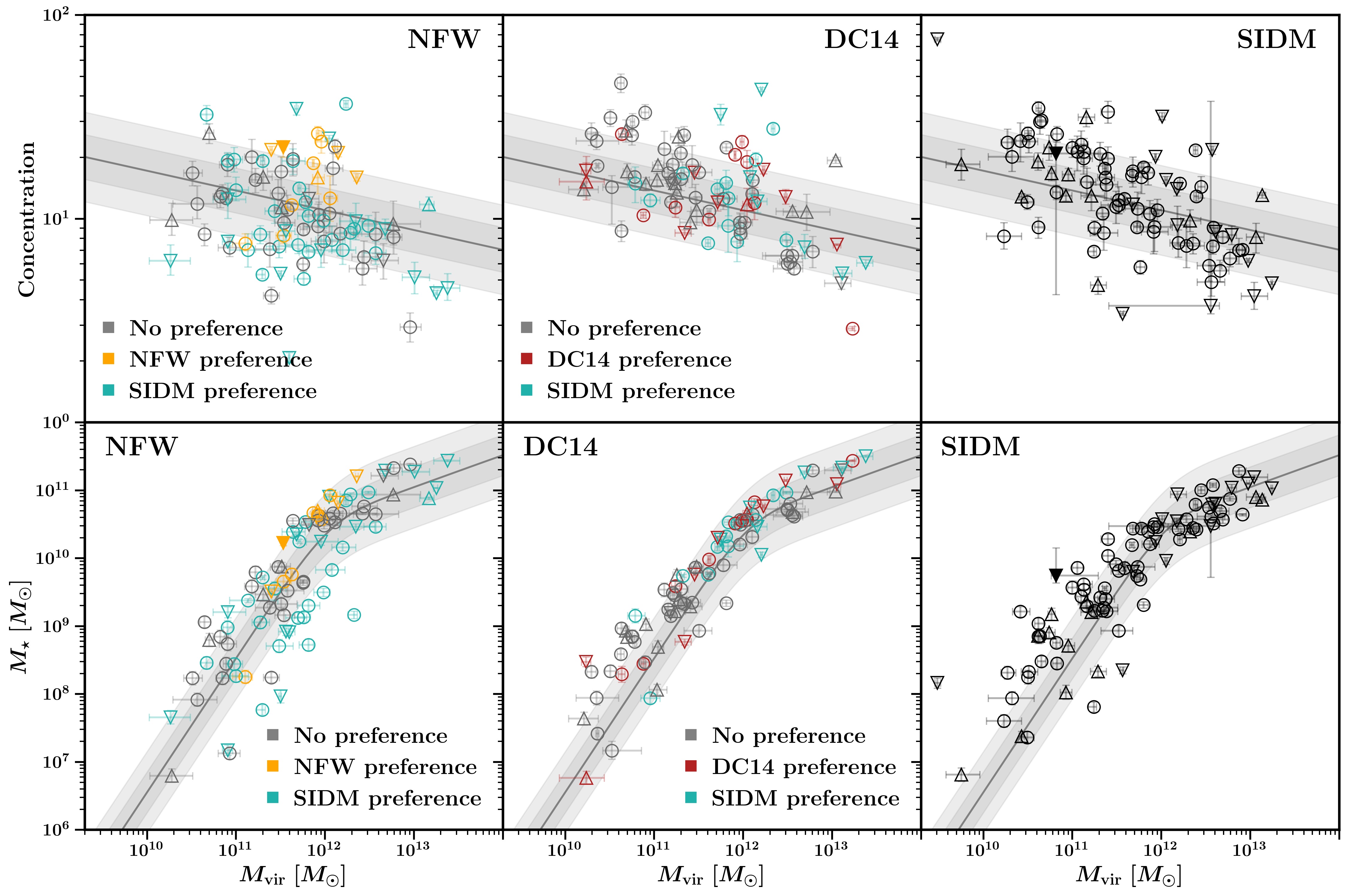}
   \caption{\emph{(Top Row)} The best-fit concentration, $c_{\rm vir}$, for each SPARC galaxy plotted as a function of its best-fit halo virial mass for the NFW~(left), DC14~(middle), and SIDM~(right) models.  The solid black line corresponds to the best-fit concentration-mass relation from~\cite{Dutton_2014}; the gray bands correspond to $\pm1$ and $2\sigma$ spread with $\sigma = 0.11$~dex.  Note that our benchmark analysis enforces this concentration-mass relation at the prior level.  \emph{(Bottom Row)} The best-fit stellar mass for each SPARC galaxy plotted as a function of its best-fit halo virial mass for the NFW~(left), DC14~(middle), and SIDM~(right) models.  The solid black line denotes the abundance matching relation from~\cite{behroozi2019}; the gray bands correspond to $\pm1$ and $2\sigma$ spread with $\sigma = 0.3$~dex. Importantly, the abundance-matching relation is not enforced at the prior level for the benchmark analysis, but it is clearly recovered by the fitting procedure.  For all panels, each point corresponds to the median parameter value, with the lines spanning the 16--84$^\text{th}$ percentiles of the posterior distribution.  A galaxy with a suspected over (under)-fitting issue is denoted by an upwards (downwards)-pointing triangle instead of a circle.  If a galaxy fails the autocorrelation test, it is denoted as a filled marker.  For each CDM panel, the aqua points denote galaxies that favor SIDM, the yellow (red) points denote galaxies that favor the NFW (DC14) model, and the gray points denote galaxies with no strong model preference ($|\Delta\text{BIC}| < 6$).
   }
   \label{fig:conc_mass}
\end{figure*}

\begin{figure*}[t] %  figure placement: here, top, bottom, or page
    \centering
    \includegraphics[width=0.9\textwidth]{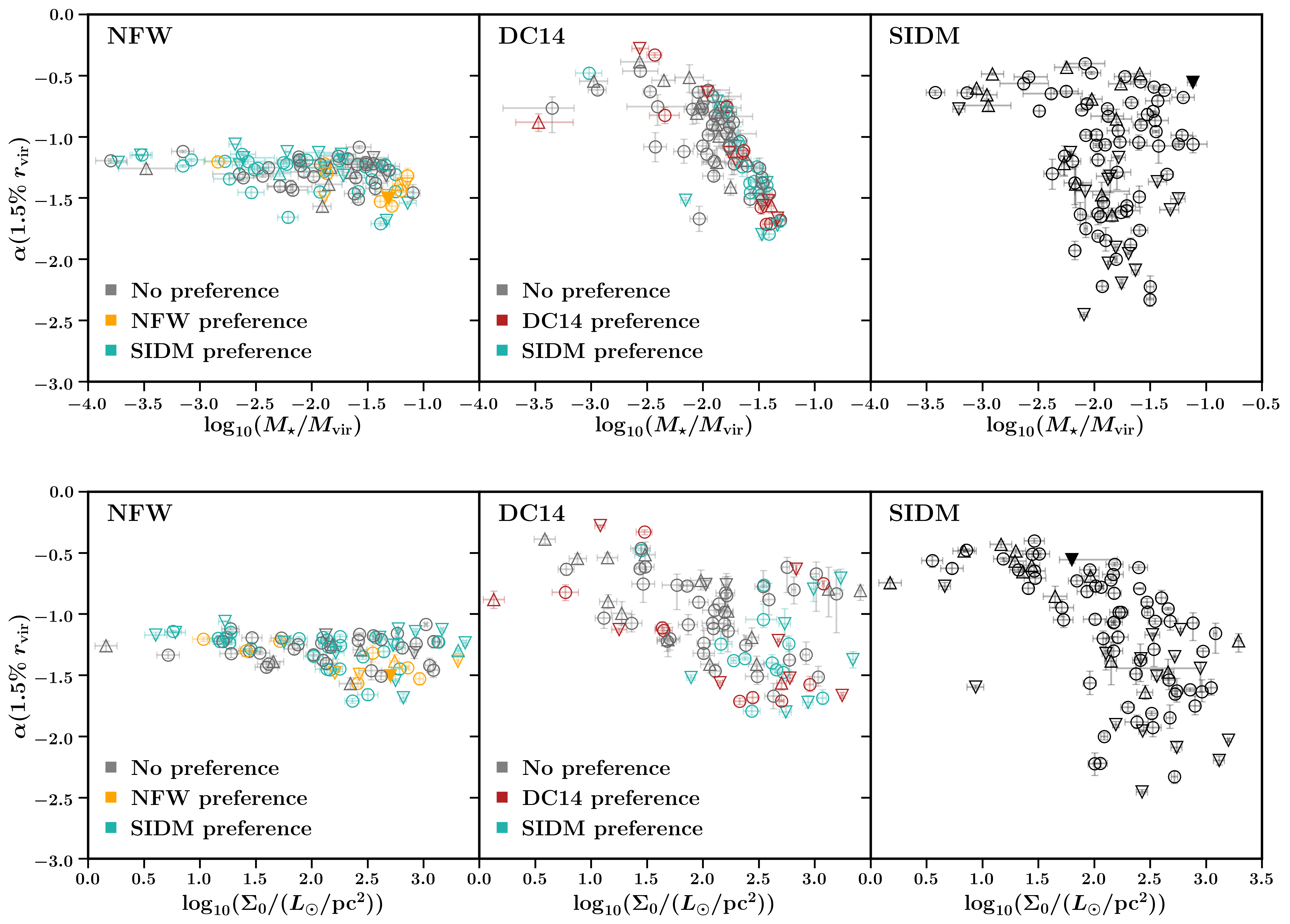} 
   \caption{The inner slope, $\alpha$, of the best-fit DM density profile as a function of $M_\star/M_\text{vir}$~(top row) and surface brightness $\log_{10}\Sigma_0$~(bottom row).  The inner slope is evaluated at 1.5\% of the virial radius, $r_\text{vir}$.  Results are provided for the NFW~(left), DC14~(middle), and SIDM~(right) models.  When the NFW model is used, the inner slope is consistently cuspy.  In contrast, the DC14 and SIDM models have enough flexibility to yield cored profiles for low-mass galaxies and cuspier profiles for higher-mass systems.  A similar transition from cored to cuspy profiles is observed moving from low to high surface brightness systems.  Compared to DC14, the SIDM model yields a larger spread in inner slopes for the high surface brightness (also, high-mass) galaxies.  The plot formatting is the same as in Fig.~\ref{fig:conc_mass}.}
   \label{fig:innerslope}
\end{figure*}

For the rotation curve analysis, it is important to identify galaxies where a particular model severely over- or under-fits the data.  These cases can be identified if the distribution of residuals is markedly different from a Gaussian with $\mu = 0$ and $\sigma^2 = 1$.  To flag these instances, we use the Kolmogorov-Smirnov~(KS) test, which allows one to determine whether an observed distribution is drawn from some reference probability distribution function.  We also use the Shapiro-Wilk~(SW) test, which is sensitive to the normality of the observed distribution.  Whether the KS or SW test is better able to distinguish the observed distribution of normalized residuals from the Gaussian distribution depends on a variety of factors, including the number of measured data points and the dispersion of $R_i$ values.  As a result, we identify any galaxy with a potential over/under-fitting issue if it has a KS or SW p-value of $< 0.05$.   We require that all galaxies have at least 10 data points in their rotation curve to ensure that these p-values are robust.

Fig.~\ref{fig:overfitting} provides a concrete example of these considerations.  The left panel shows the rotation curve data for F583-1. The corresponding best-fit SIDM, DC14 and NFW models are also provided. The right panel shows the corresponding residuals, with the black curve indicating the target normal distribution.  Both the DC14 and SIDM models over-fit the data as the distribution of residuals has $\sigma^2 < 1$; the NFW residuals do not suffer this issue.  This is also visually apparent from the rotation curve in the left panel, where it is clear that the aqua and red lines pass almost perfectly through each data point, despite the size of the error bars on the data.

Of the 90 galaxies studied, 7 are flagged as over-fit for the NFW model, 16 for DC14, and 13 for SIDM. Another 20 are flagged as under-fit for the NFW model, 16 for DC14 and 16 for SIDM. Between all three models, there are 4~(11) galaxies that are jointly classified as over-fit~(under-fit).  Between the DC14 and SIDM models alone, there are 10~(13) galaxies that are jointly classified as over-fit~(under-fit).  In general, we find that the under-fitting issues for the DC14 and SIDM models are overwhelmingly with the high surface brightness galaxies, and the over-fitting issues are with the low surface brightness galaxies.  For the NFW scenario, we do not find as clear a pattern for the under-fit and over-fit galaxies.

\section{\label{results:models}Model Comparison}

Before comparing the different DM models, one should verify that all three scenarios considered (NFW, DC14, and SIDM) recover well-understood concentration-mass and stellar mass-halo mass~(SMHM) relations.  Fig.~\ref{fig:conc_mass} summarizes our results for the NFW, DC14 and SIDM scenarios, shown in the left, middle, and right columns, respectively.  If a galaxy is indicated by a triangle, then it failed the goodness-of-fit criteria; an upwards-pointing triangle corresponds to an over-fit model and a downwards-pointing triangle corresponds to an under-fit model.  If a galaxy is denoted by a filled marker, then it failed the autocorrelation test.

In the top row, the concentration, $c_{\rm vir}$, is plotted as a function of halo virial mass, $M_{\rm vir}$.  The best-fit Planck concentration-mass relation from~\cite{Dutton_2014} is indicated by the black line, with the gray bands showing the $\pm1$ and $2\sigma$ spread with $\sigma = 0.11$~dex.  This relation was enforced with a log-normal prior (see Tab.~\ref{tab:priors}). The recovered concentrations correspond reasonably well with the theoretical expectation for all three cases, although with larger scatter.  In particular, we find that approximately 37\%~(43\%)~(36\%) of the galaxies fall inside the $\pm1\sigma$ band for NFW~(DC14)~(SIDM); the corresponding fraction of galaxies within the $\pm2\sigma$ band is 66\%~(74\%)~(74\%).

The recovered SMHM for each model is provided in the bottom row of Fig.~\ref{fig:conc_mass}.  The best-fit distribution from~\cite{behroozi2019} is also shown for comparison.  The gray bands correspond to an approximate $\pm1$ and $2\sigma$ spread with $\sigma = 0.3$~dex.\newfootnote{While the best-fit scatter in stellar mass depends on the halo mass~\citep{behroozi2019}, this approximation is a good proxy for the range of halo masses relevant to the SPARC catalog.}
It should be stressed that the SMHM relation has not been included at the prior level. Indeed, the only requirement on the stellar mass in our benchmark analysis comes from the conservative prior $M_{\rm b, tot}/M_{\rm vir} \leq 0.2$.  The recovered SMHM relation roughly tracks the expectation for all three models, although the DC14 case yields a dispersion that is most closely aligned with the expectation. In this case, we find that approximately 36\%~(52\%)~(51\%) of the galaxies fall inside the $\pm1\sigma$ band for NFW~(DC14)~(SIDM); the corresponding fraction of galaxies within the $\pm2\sigma$ band is 70\%~(81\%)~(69\%).

\begin{figure*}[t] %  figure placement: here, top, bottom, or page
    \centering
    \includegraphics[width=0.9\textwidth]{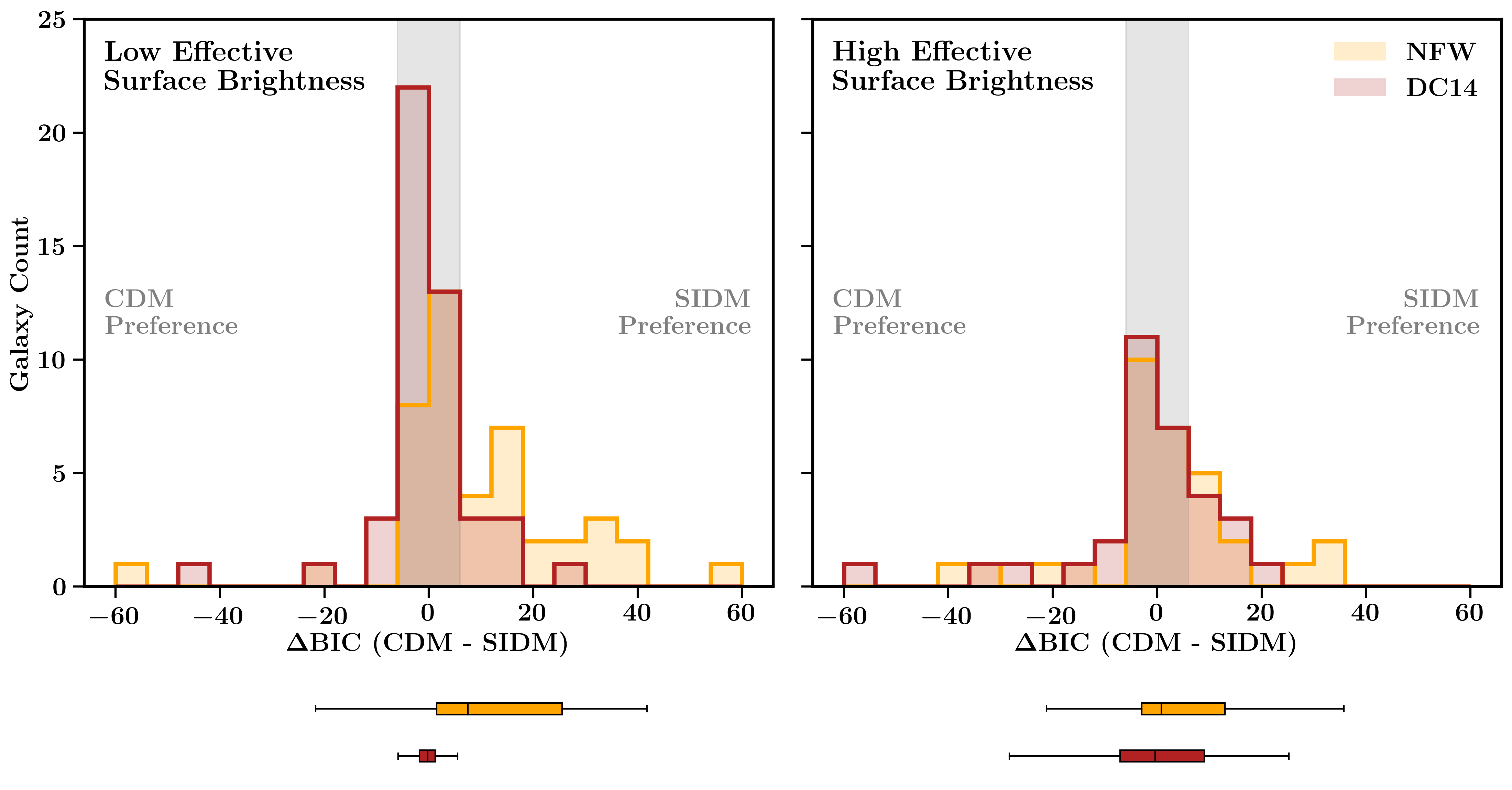}
   \caption{Distribution of \dBIC values for all SPARC galaxies with $\log_{10} \left(\Sigma_{\rm eff}/\left(L_\odot/\text{pc}^2\right)\right) < 2.5$~(left) and $>2.5$~(right).  Positive values indicate a preference for SIDM over NFW~(orange) and DC14~(red).  The vertical gray band indicates regions of \dBIC where the preference for either model is not  strong.  For each panel, the colored box and whisker plots summarize the spread of the corresponding histogram.  The box extends from the first to the third quartile of $\Delta$BIC values, with the median indicated by the vertical black line.  The whiskers extend a factor of 1.5 times the interquartile range.  In general, the SIDM model provides a better fit to the rotation curve data than the NFW model, especially for low effective surface brightness galaxies.  Additionally, there is no significant preference for either the SIDM or DC14 models for either the low or high effective surface brightness galaxies.}
   \label{fig:BIChistogram}
\end{figure*}

For each galaxy in the sample, we compare the Bayesian Information Criterion~(BIC) obtained for each model and consider
\begin{equation}
\Delta\text{BIC}  = \text{BIC(model 1)} - \text{BIC(model 2)} \, .
\end{equation}
When \dBIC$>6$, this indicates a strong preference for Model~2 over Model~1 and vice versa when \dBIC$<-6$~\citep{kass1995bf}.  For all other cases, we consider the two models statistically indistinguishable. In Fig.~\ref{fig:conc_mass}, we illustrate the preference for SIDM against either NFW or DC14. Each point in the left-most and central panels is colored by its $\Delta\text{BIC}$ value with aqua indicating a strong preference for SIDM, yellow indicating a strong preference for NFW, and red indicating a strong preference for DC14. Gray indicates no preference between SIDM and the CDM model being considered.  

The SIDM model is preferred over NFW for many of the galaxies over the entire range of masses. In contrast, most galaxies in the low-mass range show no strong preference between SIDM and DC14.  At high masses, there are approximately an equal number of galaxies with preference for either DC14 or SIDM. This suggests that both the SIDM and DC14 models provide enough flexibility to capture the diversity of the rotation curves in the lowest-mass SPARC systems, especially compared to NFW.  Of galaxies with larger virial masses, some exhibit a strong preference for one model over another, but there is no systematic preference for any of the models when the galaxies are considered in aggregate.

Fig.~\ref{fig:innerslope} illustrates how the inner slope of the best-fit density profile, $\alpha$, depends on $M_\star/M_\text{vir}$~(top row) and on surface brightness $\log_{10} \Sigma_0$~(bottom row).  As in Fig.~\ref{fig:conc_mass}, the $\Delta$BIC value for each galaxy is indicated by the color of the point; if a data point is marked as a triangle, it indicates that the galaxy failed the goodness-of-fit criteria.  The inner slope is evaluated at 1.5\% of the virial radius.  In the NFW case, the best-fit density profiles are consistently cuspy, as is enforced by the shape of the density profile, with $\alpha \in [-1.5, -1]$ across the complete galaxy sample.  For the DC14 case, the low surface brightness galaxies exhibit coring, with best-fit inner slopes closer to $\alpha \sim 0$.  Towards higher-surface brightness galaxies, the inner-slopes decrease in value and the density profiles become cuspier, similar to the expectation for NFW profiles.  As for the DC14 model, the SIDM model results in cored low surface brightness galaxies. 
At the high surface brightness range, the SIDM fits result in a wide range of inner slopes, spanning $\alpha \in [-3, -0.5]$.\newfootnote{Note that our analytic model for the SIDM density profile does not account for adiabatic contraction of the halo~\citep{1986ApJ...301...27B, Gnedin:2004cx}, which can have an effect for $\log_{10}M_\star/M_{\rm vir} \gtrsim -2$ and increase the inner density of the DM profile. The inclusion of adiabatic contraction effects in SIDM halos will be explored in future work (see~\cite{jiang_forwardcite}).}

Fig.~\ref{fig:BIChistogram} shows the histogram of \dBIC values in the low and high \emph{effective} surface brightness regimes, comparing either NFW or DC14 to SIDM (the effective surface brightness is reported in the SPARC catalog). These histograms underscore the need of performing such an analysis over a large ensemble of galaxies.  Indeed, the galaxy-to-galaxy scatter in the \dBIC values can be quite substantial, and it is thus important to look at the trends of the ensemble.  

In the low effective surface brightness regime with $\log_{10} \left(\Sigma_{\rm eff}/\left(L_\odot/\text{pc}^2\right)\right) < 2.5$, SIDM is preferred over NFW with the 25-50-75$^{\rm th}$ percentiles of the $\Delta$BIC values being 1.39, 6.81, and 23.24. In contrast, there is no significant preference for SIDM over DC14, with the 25-50-75$^{\rm th}$ percentiles of the $\Delta$BIC values being $-1.63, -0.22,$ and $1.10$.  
In the high effective surface brightness regime, the preference for the SIDM model over NFW continues, although it is much weaker.  In this case, the 25-50-75$^{\rm th}$ percentiles of the $\Delta$BIC values are $-2.69, 0.71,$ and $11.79$. There continues to be no significant preference for SIDM over DC14 in this regime, with the 25-50-75$^{\rm th}$ percentiles of the $\Delta$BIC values being $-6.46, -0.37,$ and $8.23$.  These results underscore that the SIDM model is more successful at fitting the rotation curves than the NFW model, but that this behavior is ameliorated when using the DC14 model, which allows for coring due to baryonic feedback.

To better understand how our conclusions depend on some key assumptions, we perform several variations of the benchmark analysis.  For one test, we consider the effect of imposing the abundance-matching relation from~\cite{behroozi2019} at the prior level, as opposed to the prior on $M_{\rm b, tot}/M_{\rm vir}$.  Specifically, we use the best-fit parameters associated with all star-forming galaxies (row 6 in their Table~J1) and assume a constant scatter on the stellar mass of $\sigma \sim 0.3$ as a function of $M_{\rm vir}$.  For a separate test, we follow Tab.~\ref{tab:priors}, but use a linear-flat prior on the disk and bulge mass-to-light ratios: $\Upsilon_{\star, \text{disk}}, \Upsilon_{\star, \text{bul}} \in [0.1, 10]$.  For both tests, the results are highly consistent with those presented in Figs.~\ref{fig:conc_mass}--\ref{fig:BIChistogram}.  All relevant plots for these tests are available at the web address in Footnote~1.

As an outcome of our study, we are able to scrutinize individual rotation curves and their best-fit results for each model. A recent study by~\cite{2018MNRAS.473.4392S} found that NIHAO simulated galaxies were unable to reproduce SPARC rotation curves that have either very small or order-unity values for the ratio of circular velocity at 2~kpc to circular velocity at last measured radius. In Appendix~\ref{app:B}, we present our results for five galaxies that are both in our sample and are marked as outliers in~\cite{2018MNRAS.473.4392S}. We find that only one out of the five has a substantial preference for DC14 over SIDM, with the others showing no strong preference between these two models.  We also find that many of these galaxies are outliers in terms of their concentrations. The latter result could potentially explain why NIHAO was unable to reproduce galaxies of this type.

A final point worth discussion is that in this study we consider a constant cross section per unit mass of $3$ cm$^2$/g for the SIDM scenario. That said, constraints on SIDM parameter space point to a velocity-dependent cross section and therefore one should think of the constant value chosen here as an approximation to the cross section at the typical velocities of SPARC galaxies in our sample. However, we find that these galaxies actually exhibit an order of magnitude spread in $\sigma_{\rm v0}$ of around $30$--$250$~km/s, and therefore any velocity-dependent cross section cannot vary too much over that range. This requirement is likely to be in conflict with the combination of constraints arising from cluster and galaxy group relaxation~\citep{Sagunski:2020spe} and from central densities of satellite galaxies~\citep{Jiang:2021foz}. Thus, a more dedicated study of this possibility is warranted.

\section{\label{conclusions}Conclusions}

In this paper, we use the SPARC dataset to study several solutions to the galaxy diversity problem.  We focus on three specific models that capture the relevant physics of SIDM and CDM with/without baryonic feedback.  For the latter, we use the DC14 model~\citep{DiCintio:2014xia} as an example of a feedback-affected halo and an NFW model as an example of a CDM halo with no (or weak) feedback.  Improving upon the statistical methodology used in previous studies of SPARC galaxies, we perform a comprehensive model comparison to better understand the preference for the three models we consider.

Our benchmark analysis, which takes the Planck concentration-mass relation of~\cite{Dutton_2014} as a prior, roughly recovers the SMHM relation of~\cite{behroozi2019} for all three scenarios.  Both the SIDM and DC14 cases return profiles that are more cored at low surface brightness than the NFW expectation.  For high surface brightness galaxies, all three models predict cuspier profiles, although the SIDM model results in the largest spread in best-fit inner slopes.

Because of galaxy-to-galaxy variance in the model preferences, it is important to consider the trends of the sample in aggregate.  Overall, we find that the high surface brightness galaxies in the SPARC catalog have no strong statistical preference for SIDM over DC14 model and only a weak preference for SIDM over NFW.  For the low surface brightness galaxies, the SIDM model is strongly preferred over NFW, but is as good a fit to the data as the DC14 model.  We thus conclude that both SIDM as well as CDM with baryonic feedback provide adequate flexibility to explain the diversity of rotation curves in the SPARC catalog, and that the current data is not adequate to differentiate between these two scenarios.  The underlying reason for this is that the specific range of $M_\star/M_{\rm vir}$ for which coring enforced by SIDM or by baryonic feedback is most similar is also the approximate range probed by the SPARC catalog. In contrast, lower mass halos would demonstrate inherent differences between the models because $M_\star/M_{\rm vir}$ decreases with the virial mass, so coring from SIDM remains efficient even when coring from baryonic feedback terminates~\citep{2017MNRAS.472.2945R, 2020MNRAS.497.2393L}. Thus, one of the main conclusions of this study is that distinguishing between these types of physical scenarios requires rotation curve data for more low-mass objects than currently available in SPARC.

\section{Acknowledgements}
The authors gratefully acknowledge P.~Behroozi, A.~Brooks, A.~Di~Cintio, F.~Jiang, M.~Kaplinghat, H.~Liu, S.~Mishra-Sharma, M.~Moschella, and L.~Necib for fruitful conversations.  SD and AZ were supported in part by the Princeton Physics Department's undergraduate summer research program.  ML and OS are supported by the DOE under Award Number DE-SC0007968 and the Binational Science Foundation (grant No. 2018140).  This work was performed in part at the Aspen Center for Physics, which is supported by NSF grant PHY-1607611.  The work presented in this paper was performed on computatational resources managed and supported by Princeton Research Computing.  This research made extensive use of the following publicly available codes:
\texttt{IPython}~\citep{PER-GRA:2007}, 
\texttt{Jupyter}~\citep{Kluyver2016JupyterN}, \texttt{matplotlib}~\citep{Hunter:2007}, 
\texttt{NumPy}~\citep{numpy:2011},
\texttt{SciPy}~\citep{Jones:2001ab},  \texttt{emcee}~\citep{emcee}, and \texttt{Pandas}~\citep{mckinney-proc-scipy-2010}.
\clearpage

\appendix

\setcounter{equation}{0}
\setcounter{figure}{0}
\setcounter{table}{0}
\renewcommand{\theequation}{A\arabic{equation}}
\renewcommand{\thefigure}{A\arabic{figure}}
\renewcommand{\thetable}{A\arabic{table}}

\section{Corner Plots}
\label{app:corner}
This appendix provides rotation curves and corresponding corner plots for the NFW, DC14, and SIDM rotation curve fits performed for NGC0055, NGC3741, NGC3769, UGC06787, IC2574, and NGC4013---the six example galaxies spotlighted in Fig.~\ref{fig:rotationcurve}.  The posterior distributions are well-converged in all six cases.  Each galaxy also passes the autocorrelation test described in the main text (see figures provided in the \texttt{github}).

\begin{figure}[h!]
    \centering
    \includegraphics[width=0.3\textwidth]{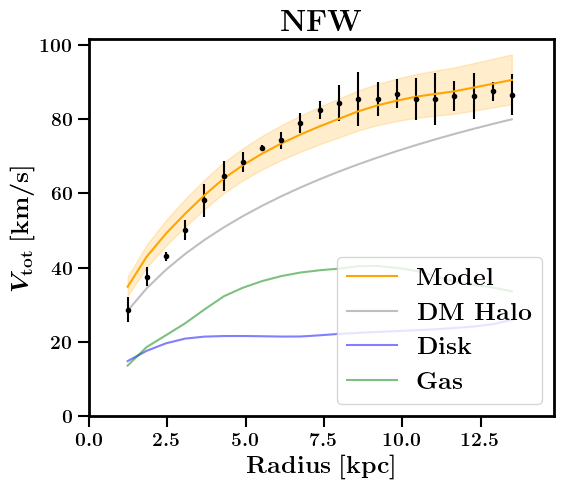}
    \includegraphics[width=0.3\textwidth]{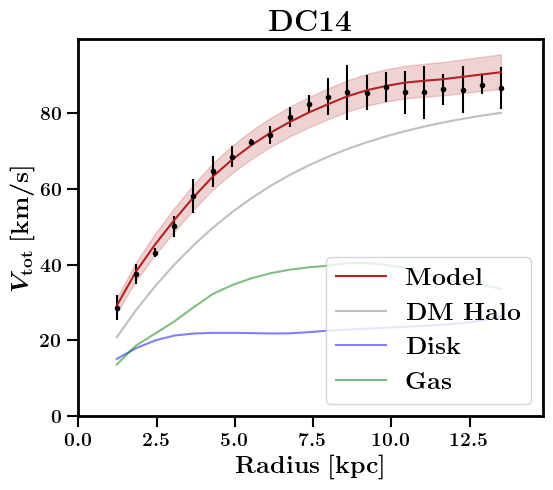}
    \includegraphics[width=0.3\textwidth]{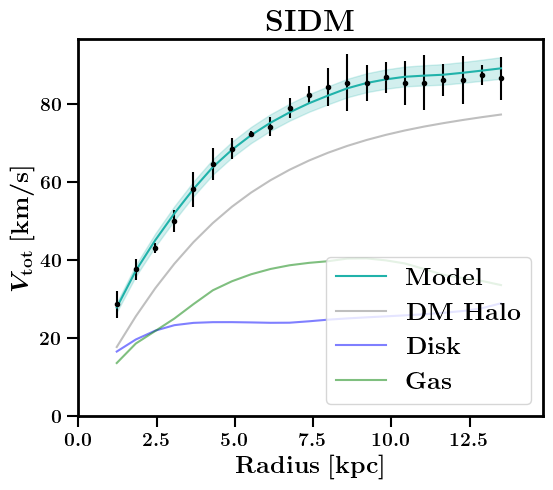}
    \includegraphics[width=0.3\textwidth]{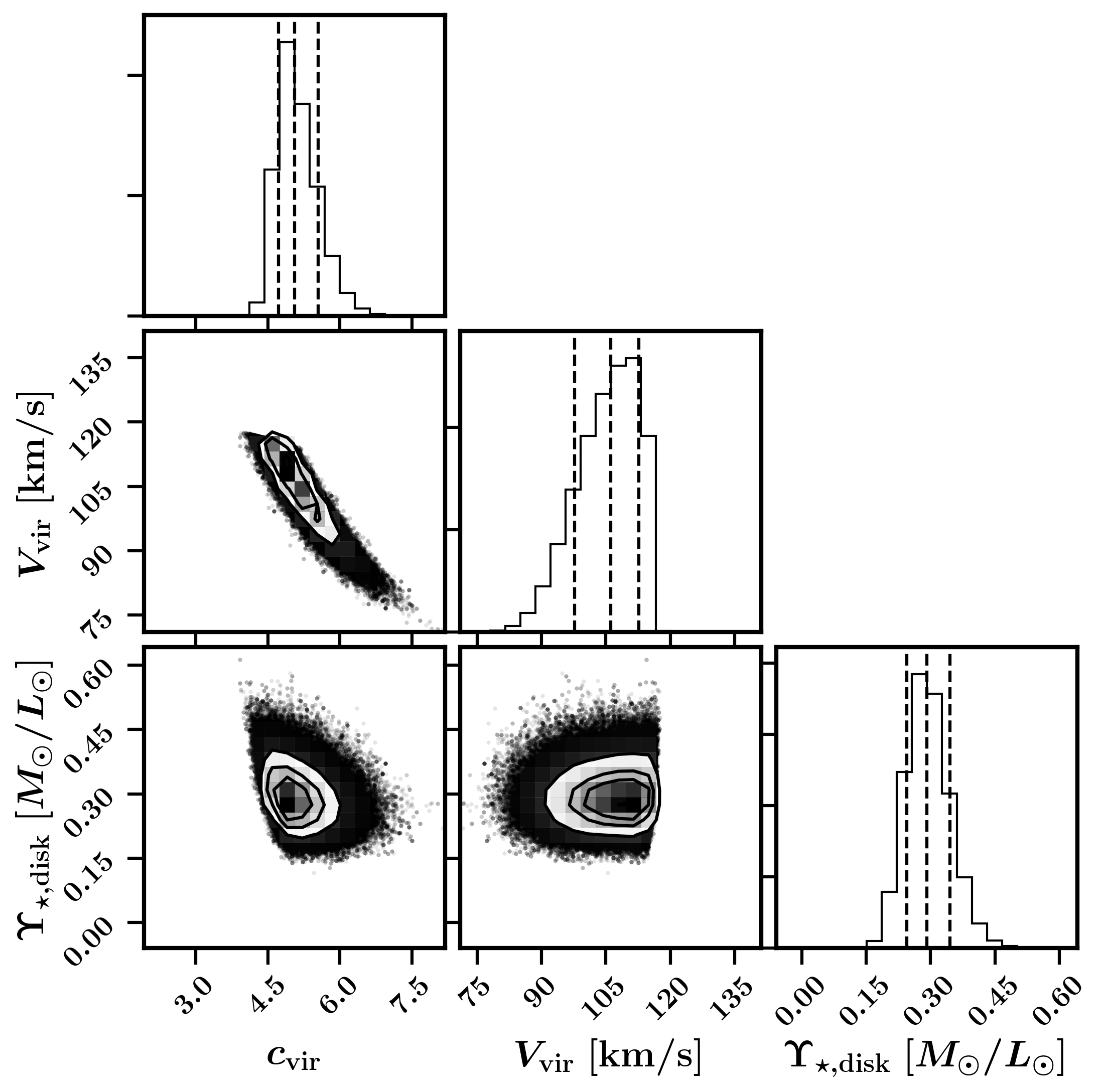}
    \includegraphics[width=0.3\textwidth]{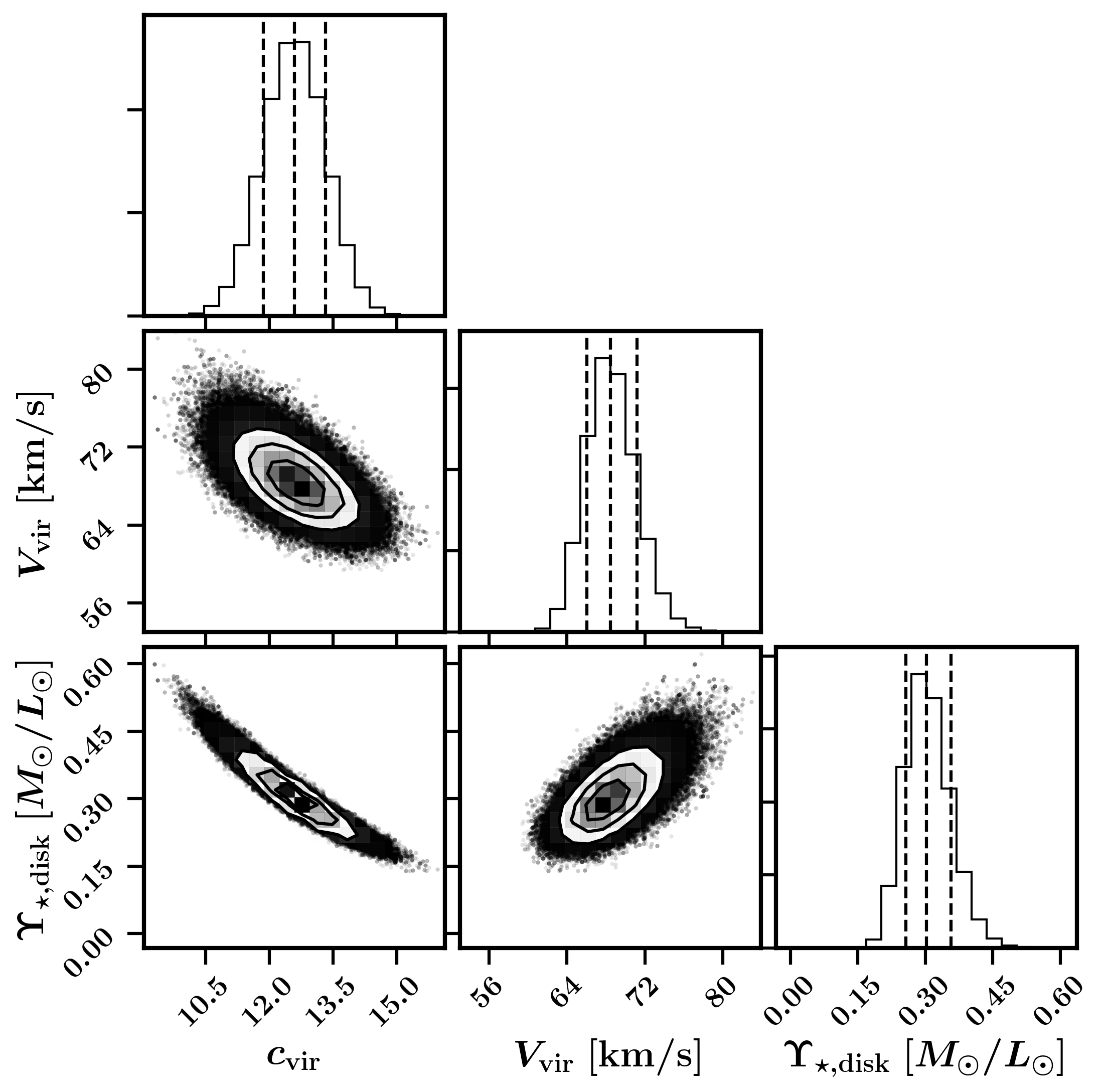}
    \includegraphics[width=0.3\textwidth]{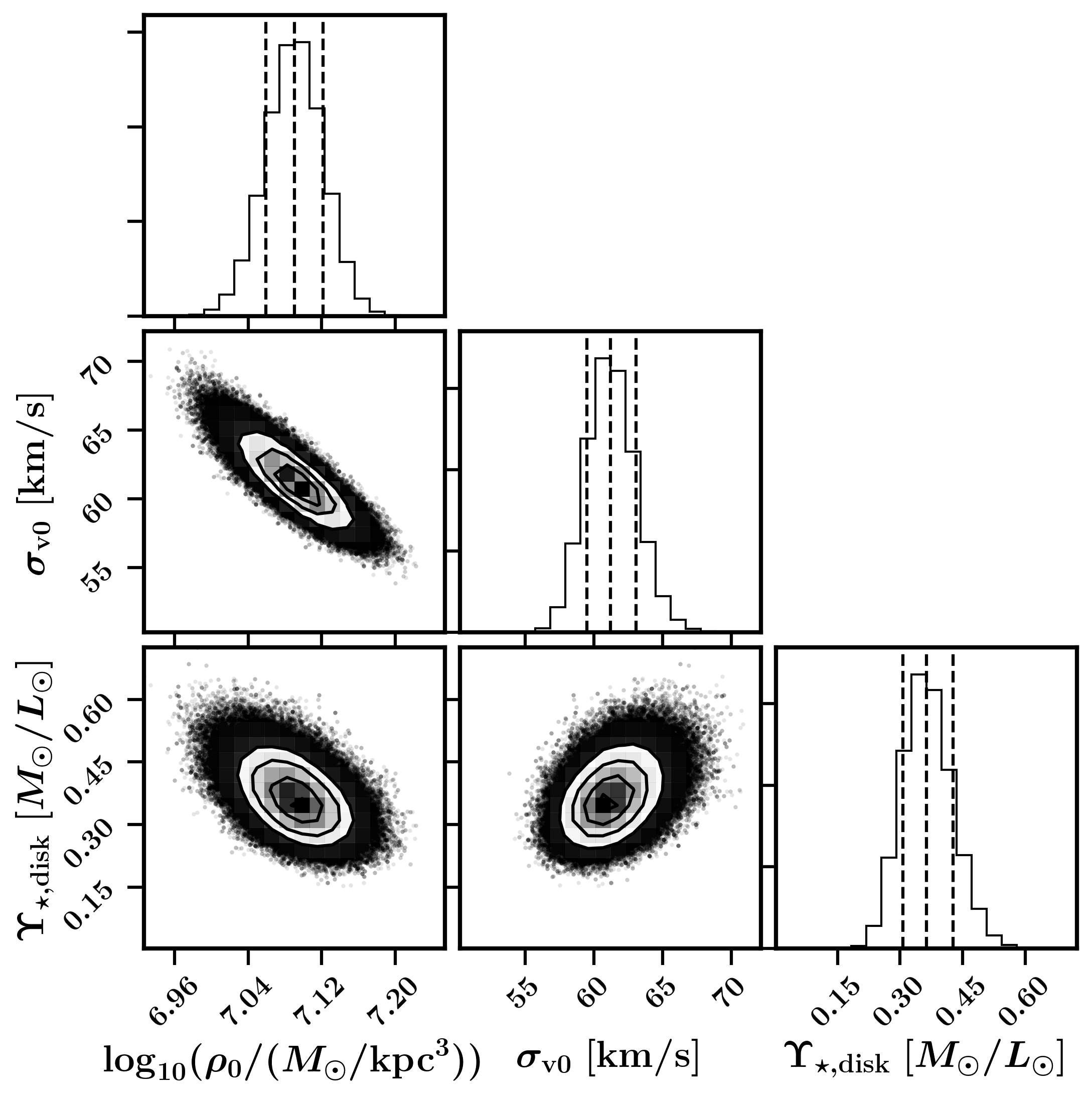}
    \caption{NFW~(left column), DC14~(middle column), and SIDM~(right column) fits to the NGC0055 rotation curve with corresponding corner plots.}
\end{figure}

\begin{figure}[h!]
    \centering
    \includegraphics[width=0.3\textwidth]{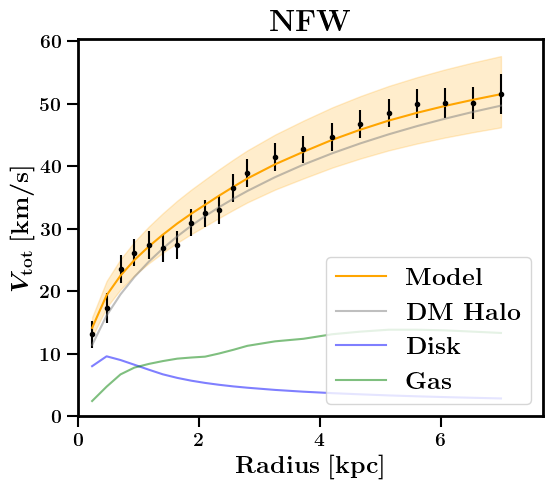}
    \includegraphics[width=0.3\textwidth]{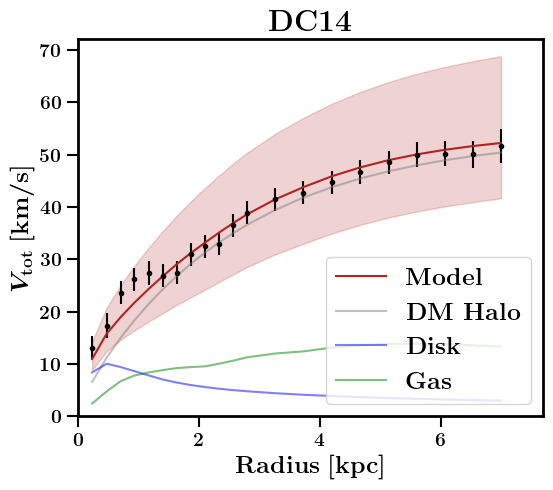}
    \includegraphics[width=0.3\textwidth]{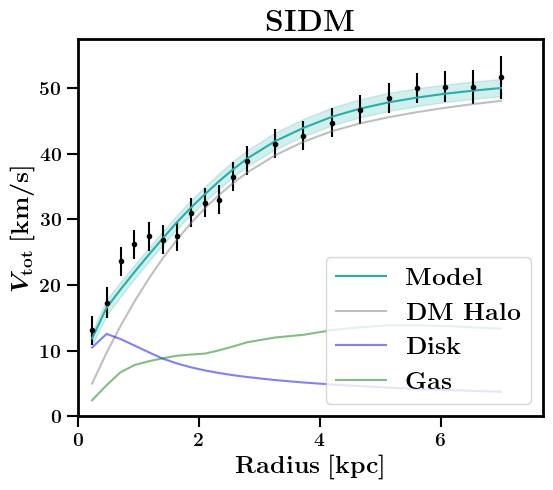}
    \includegraphics[width=0.3\textwidth]{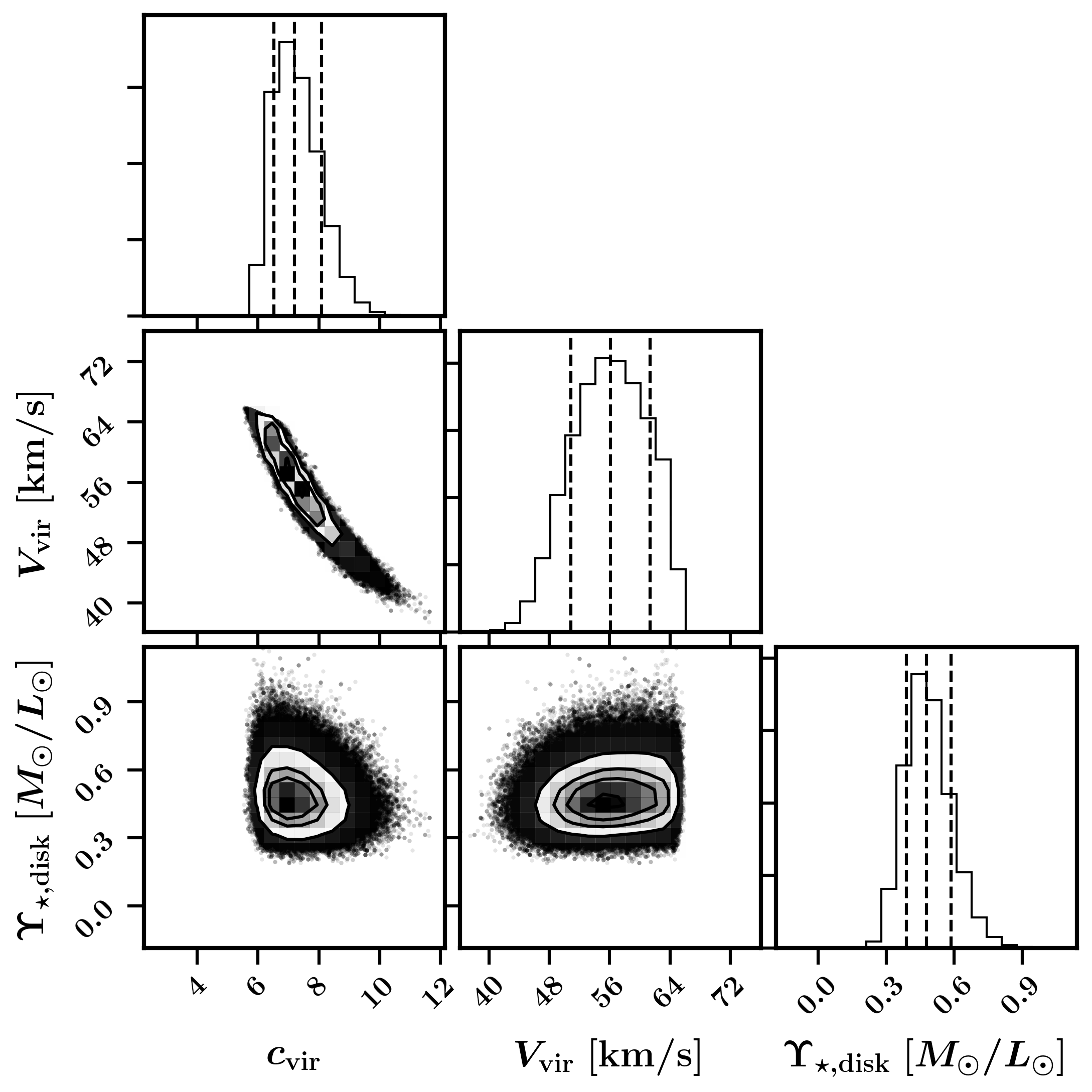}
    \includegraphics[width=0.3\textwidth]{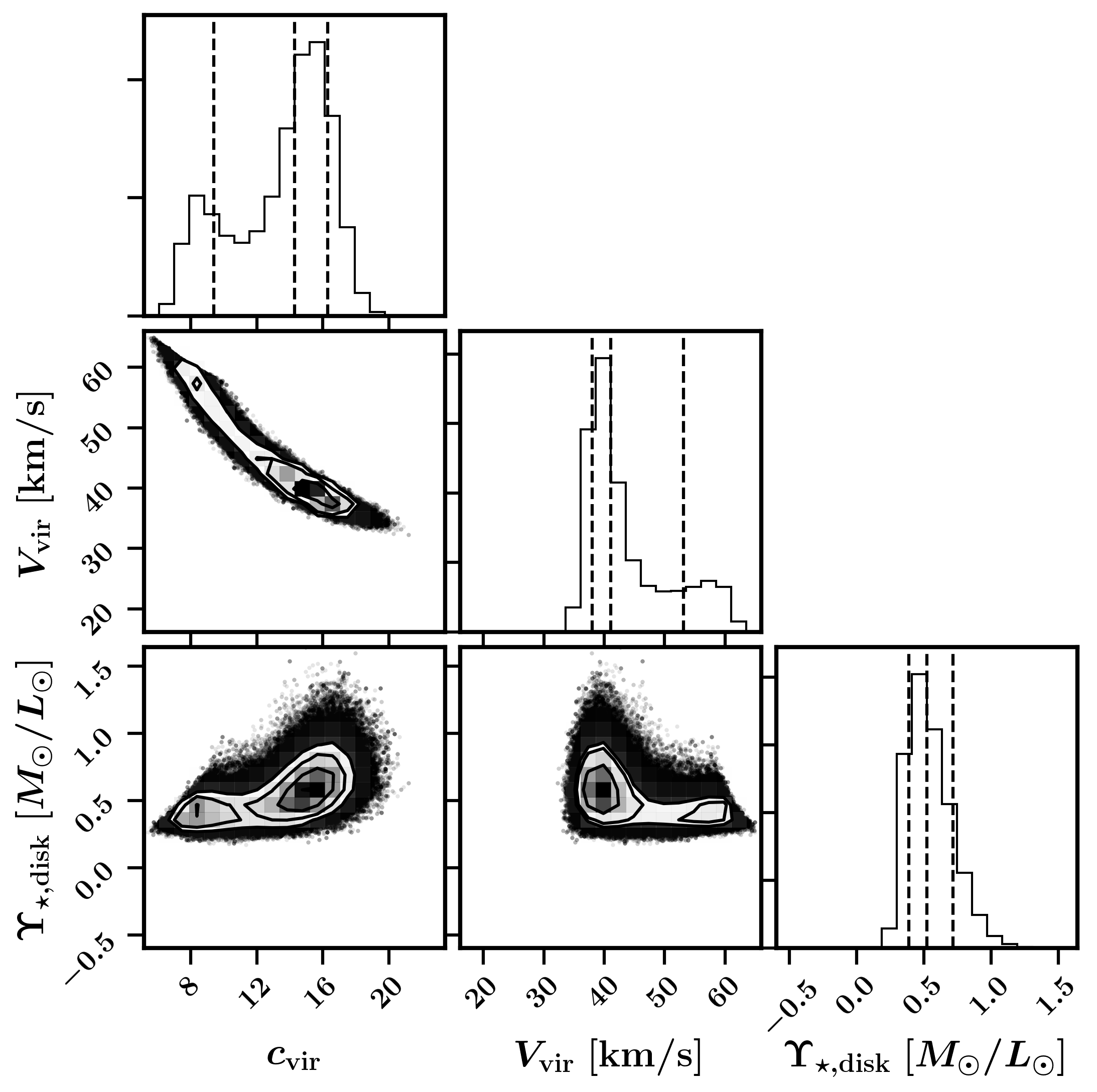}
    \includegraphics[width=0.3\textwidth]{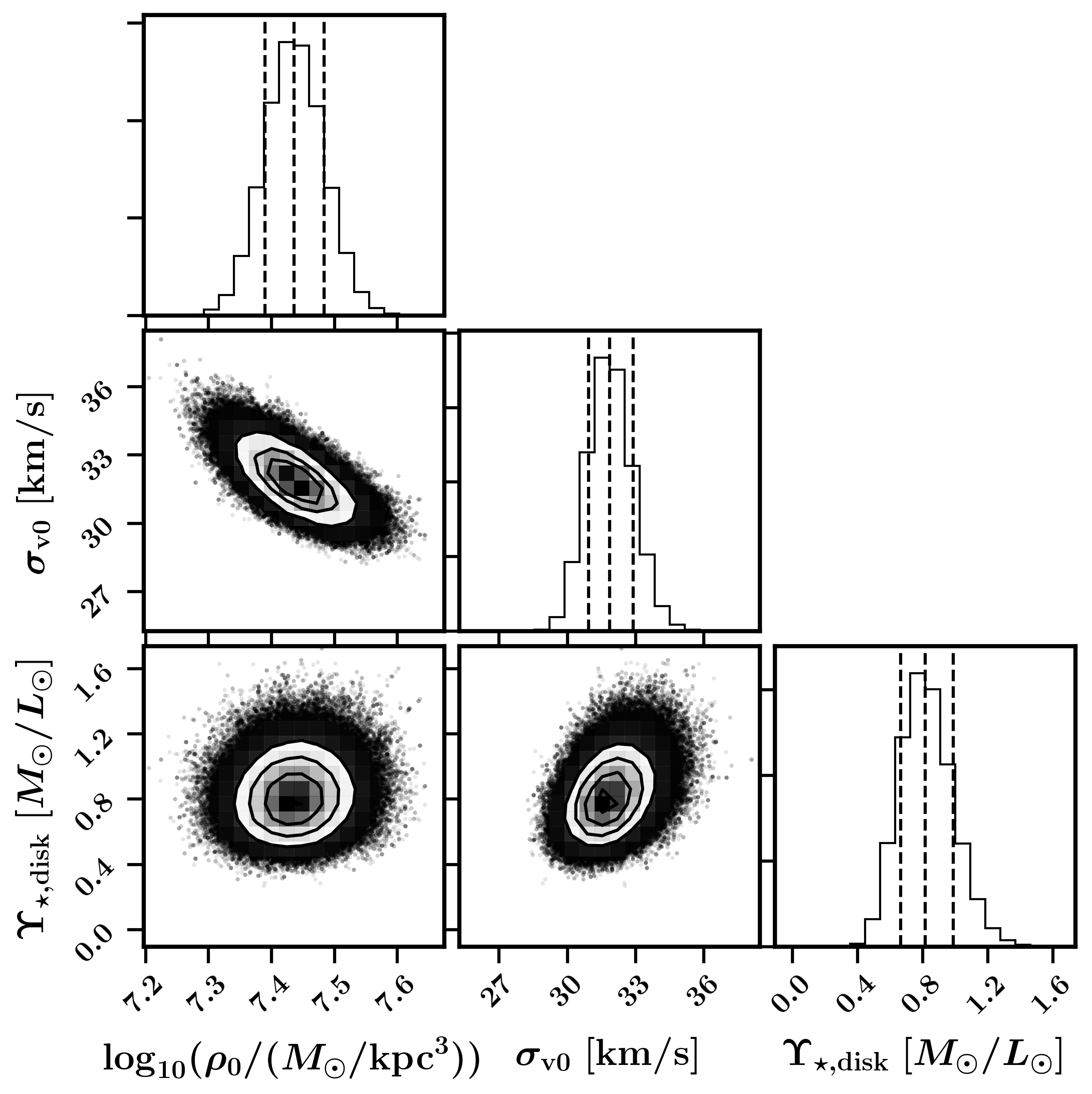}
    \caption{NFW~(left column), DC14~(middle column), and SIDM~(right column) fits to the NGC3741 rotation curve with corresponding corner plots.}
\end{figure}

\begin{figure}[h!]
    \centering
    \includegraphics[width=0.3\textwidth]{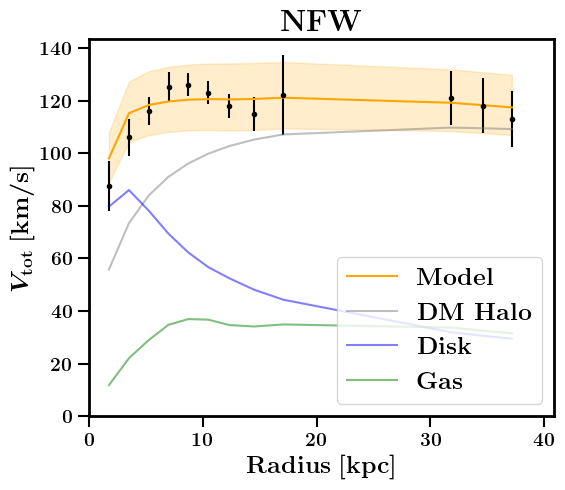}
    \includegraphics[width=0.3\textwidth]{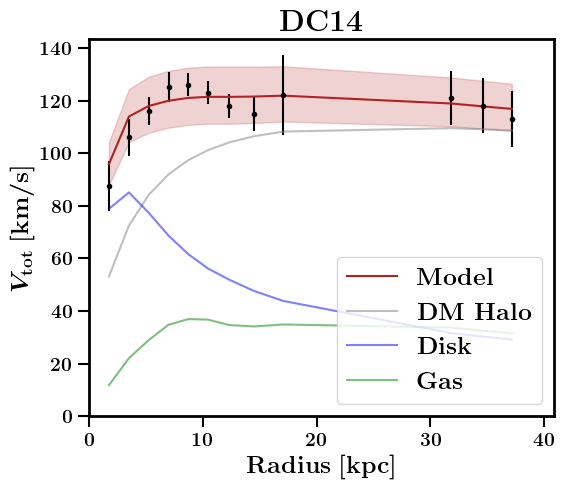}
    \includegraphics[width=0.3\textwidth]{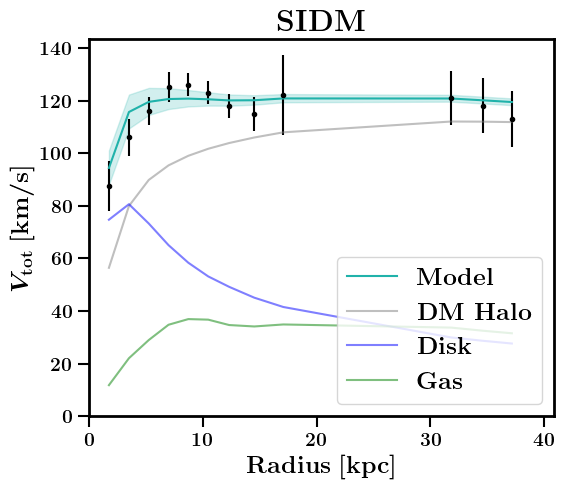}
    \includegraphics[width=0.3\textwidth]{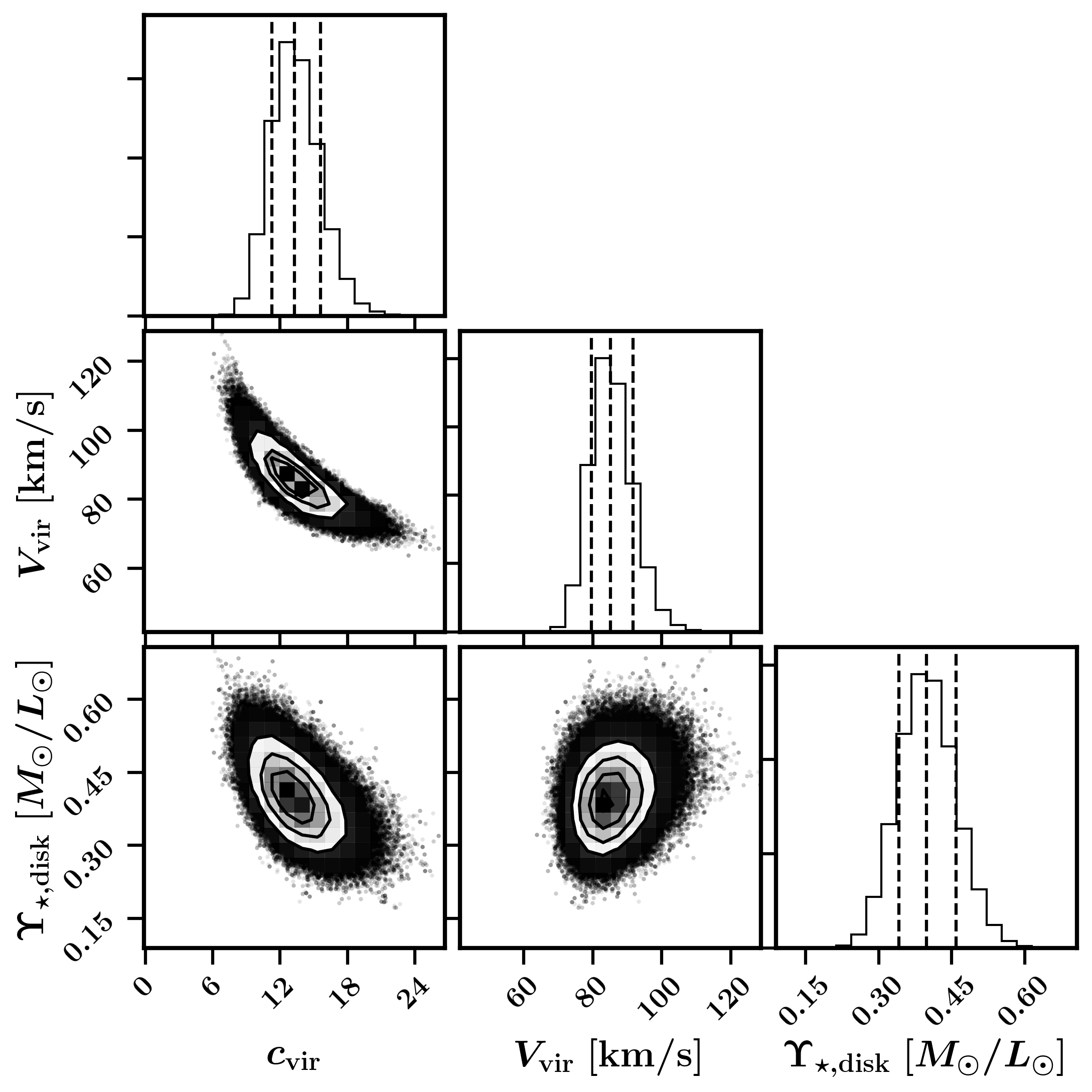}
    \includegraphics[width=0.3\textwidth]{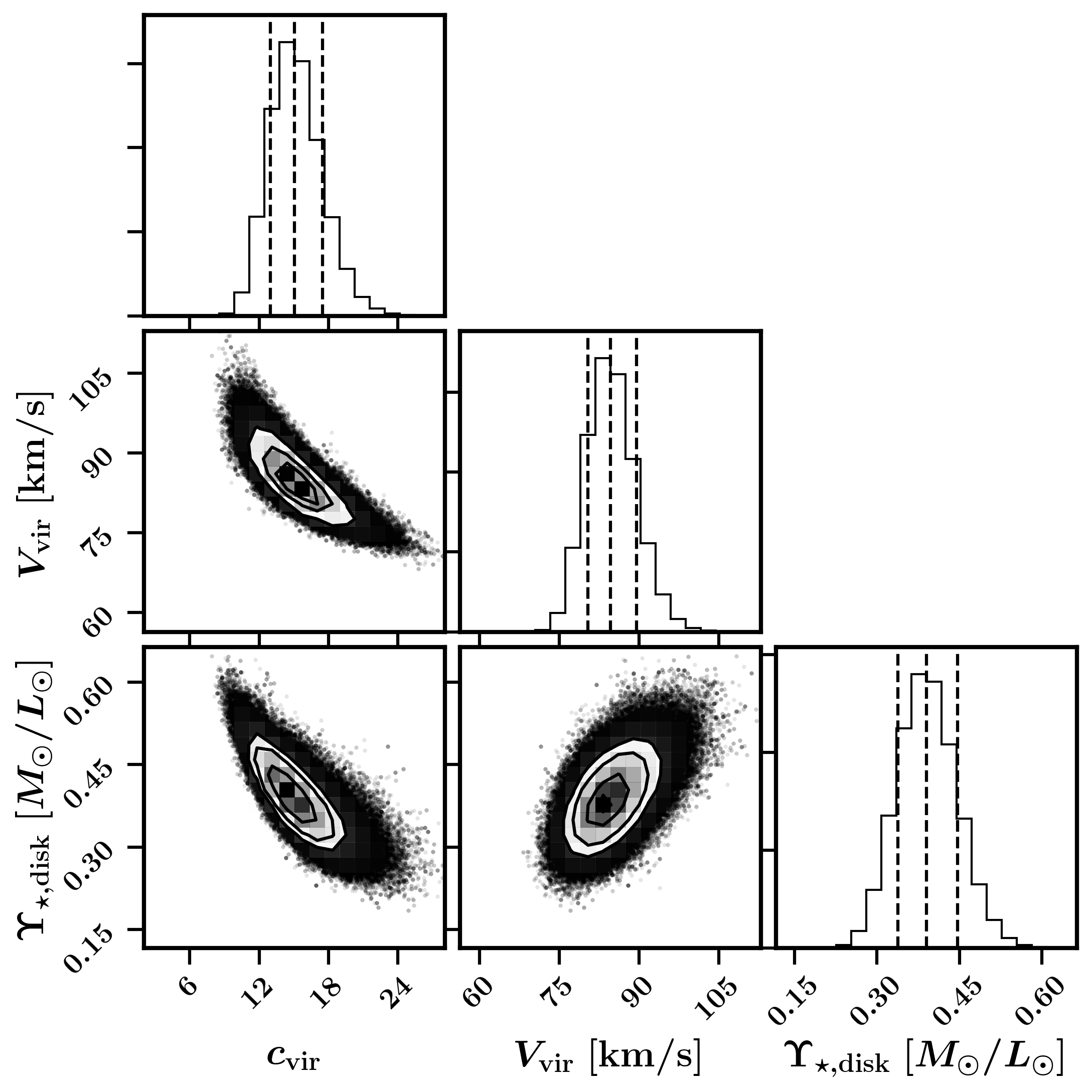}
    \includegraphics[width=0.3\textwidth]{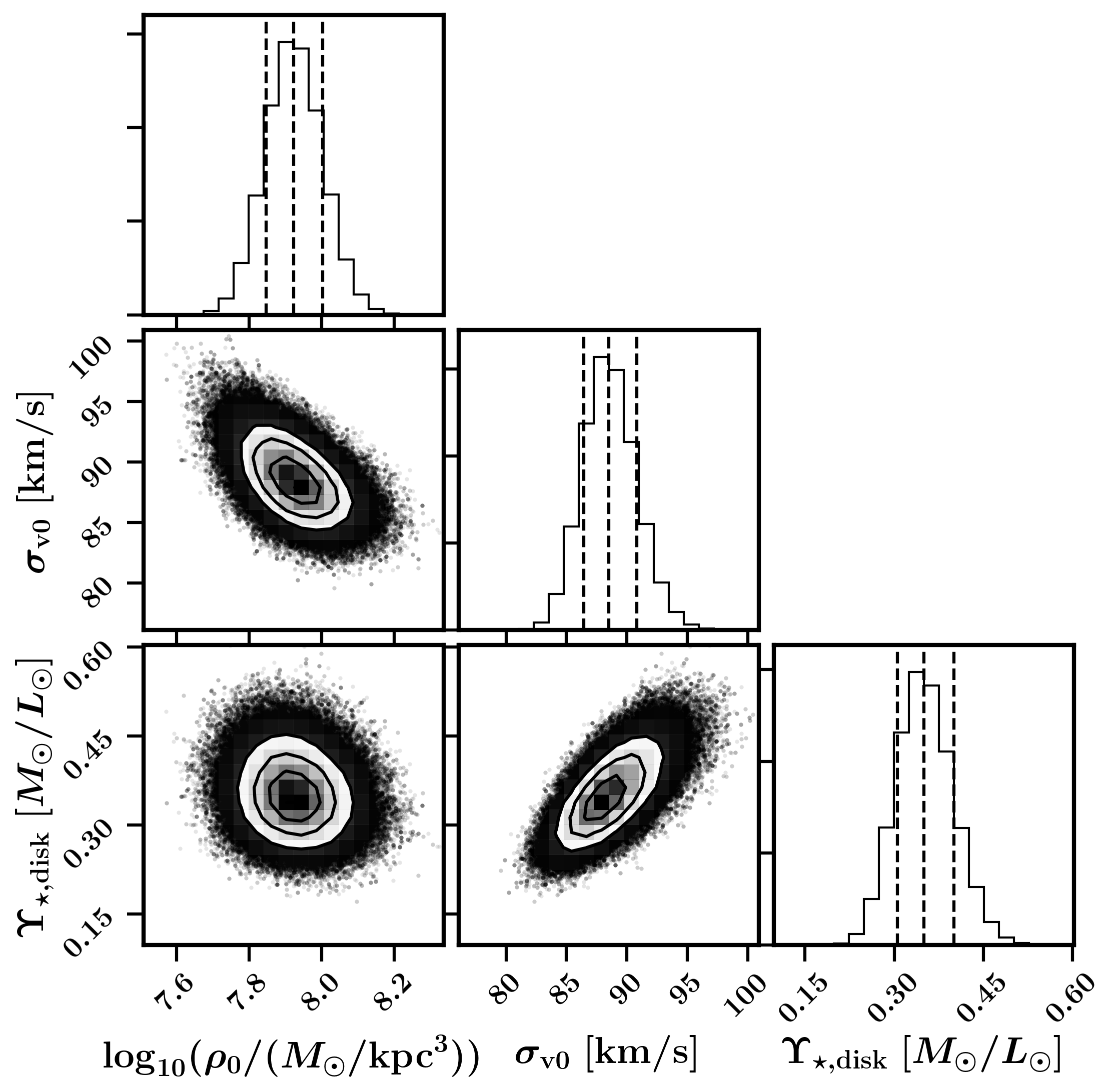}
    \caption{NFW~(left column), DC14~(middle column), and SIDM~(right column) fits to the NGC3769 rotation curve with corresponding corner plots.}
\end{figure}

\begin{figure}[h!]
    \centering
    \includegraphics[width=0.3\textwidth]{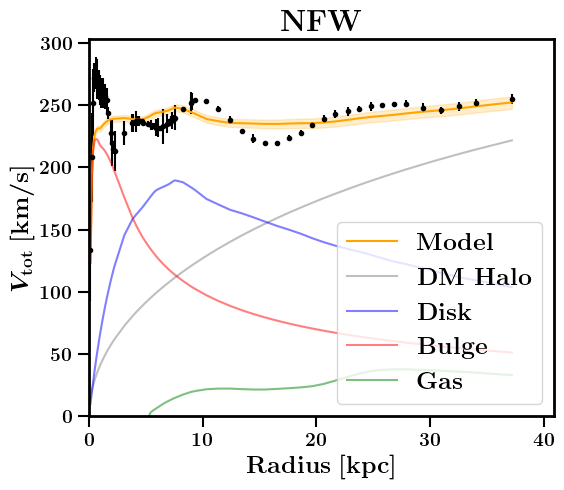}
    \includegraphics[width=0.3\textwidth]{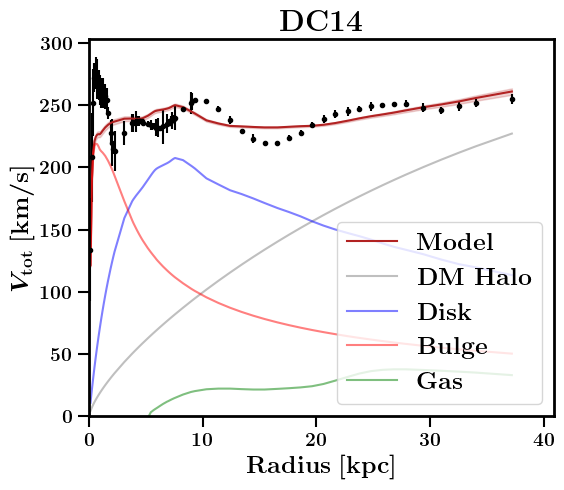}
    \includegraphics[width=0.3\textwidth]{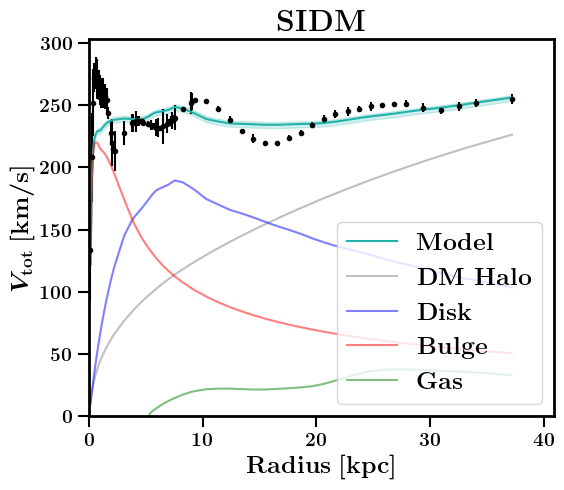}
    \includegraphics[width=0.3\textwidth]{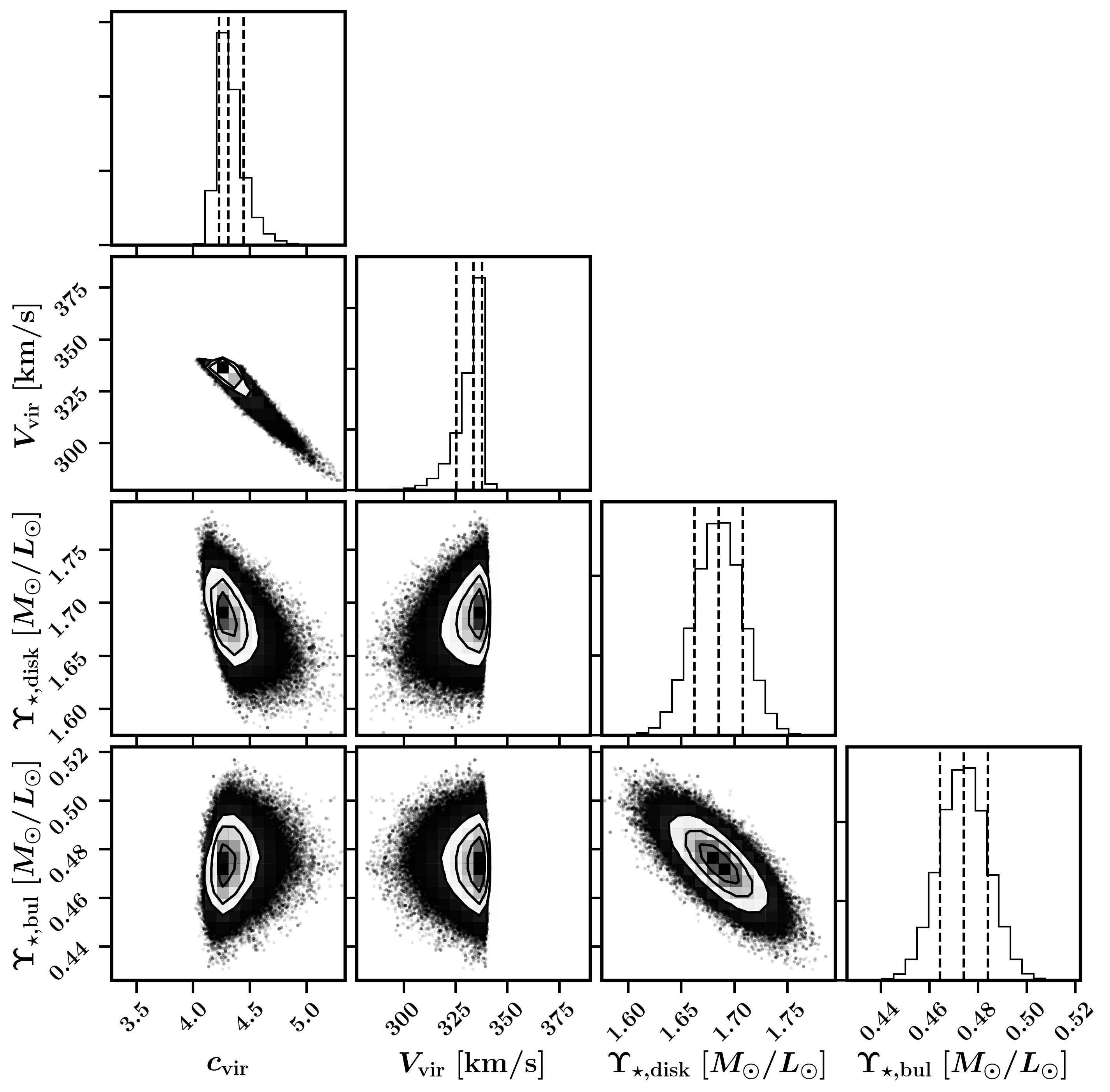}
    \includegraphics[width=0.3\textwidth]{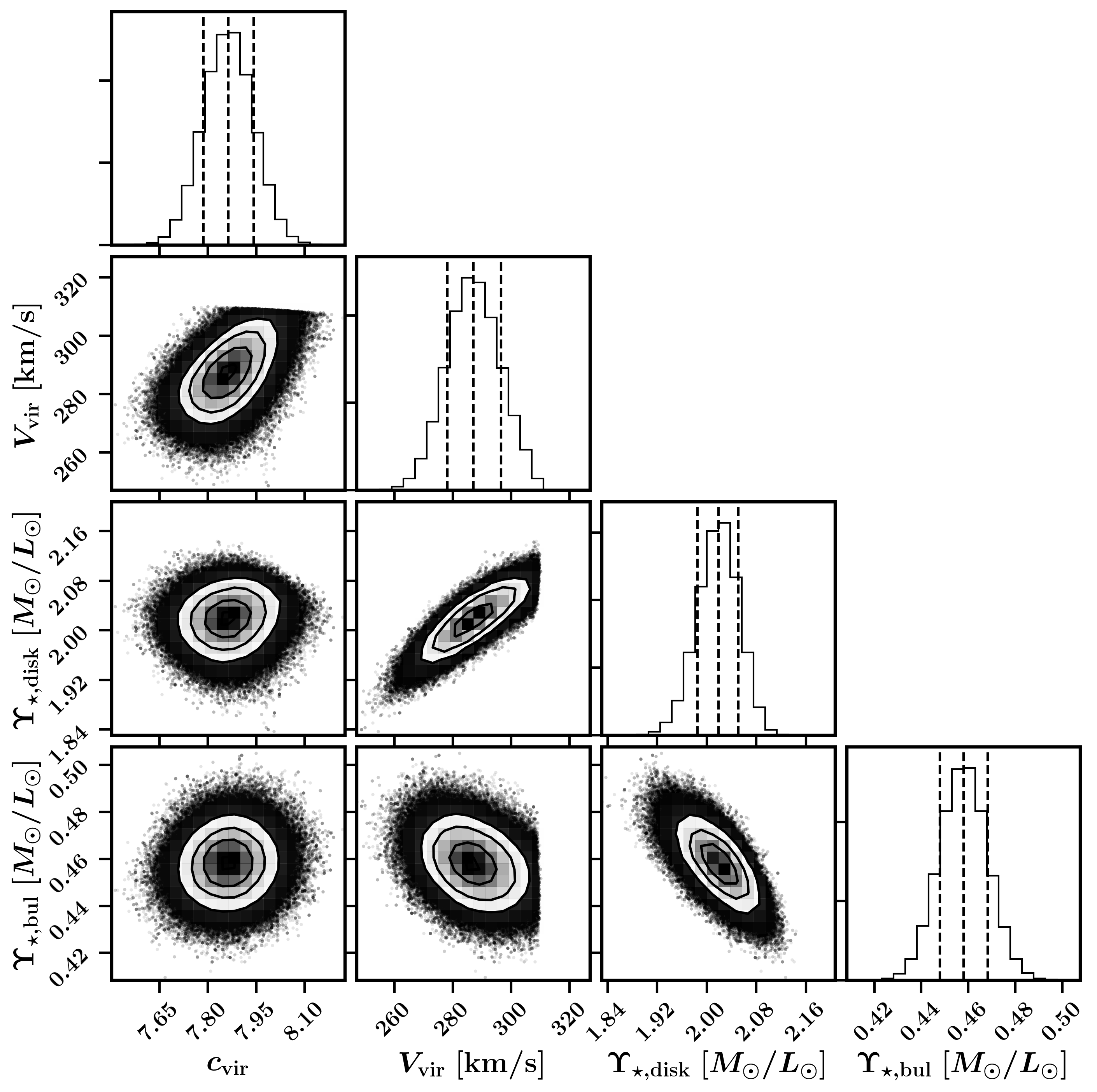}
    \includegraphics[width=0.3\textwidth]{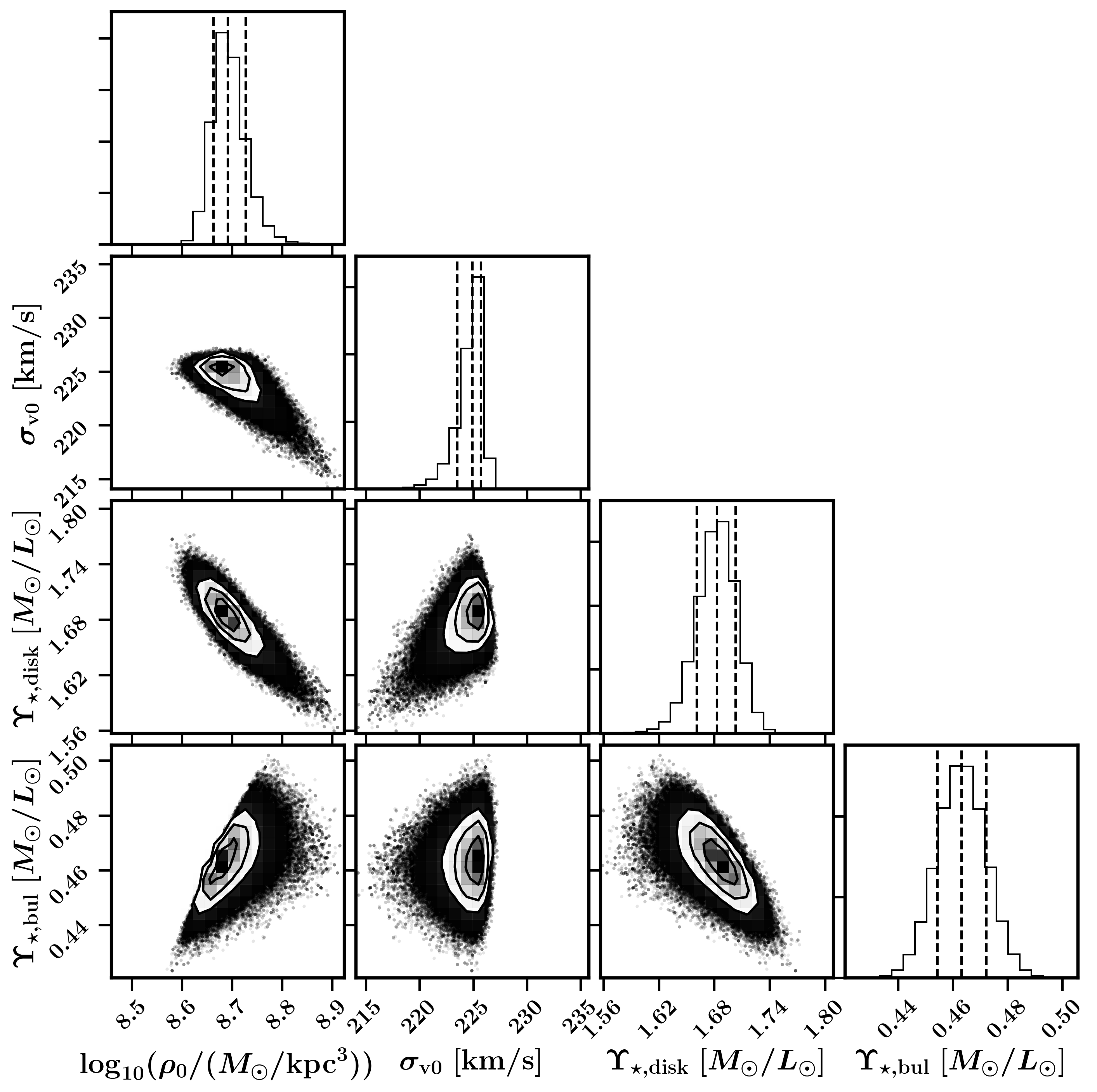}
    \caption{NFW~(left column), DC14~(middle column), and SIDM~(right column) fits to the UGC06787 rotation curve with corresponding corner plots.}
\end{figure}

\begin{figure}[h!]
    \centering
    \includegraphics[width=0.3\textwidth]{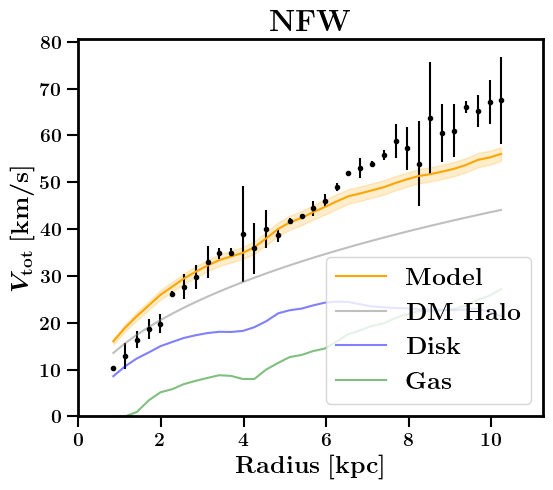}
    \includegraphics[width=0.3\textwidth]{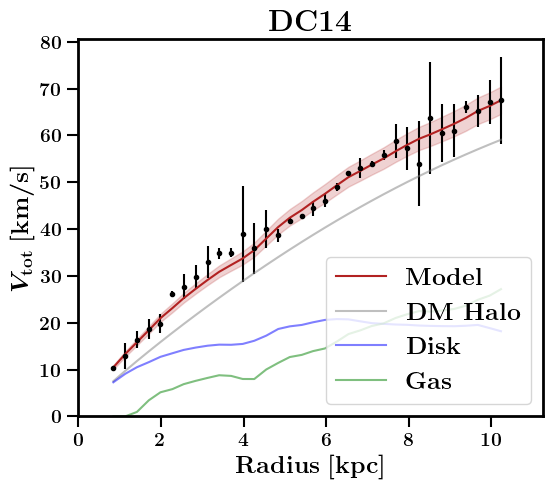}
    \includegraphics[width=0.3\textwidth]{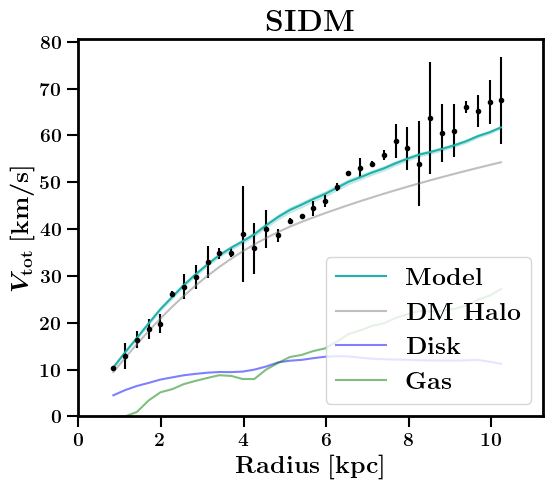}
    \includegraphics[width=0.3\textwidth]{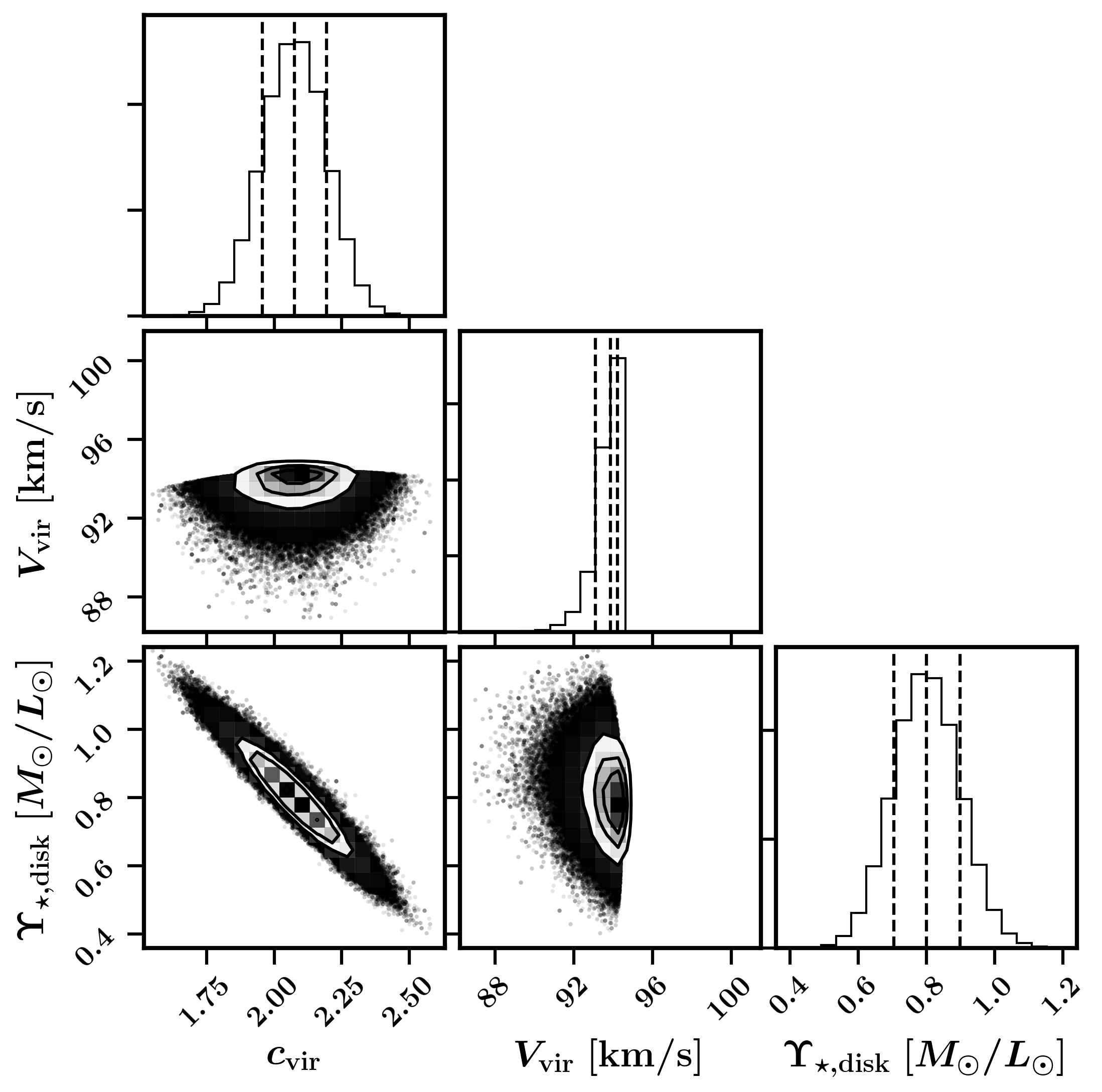}
    \includegraphics[width=0.3\textwidth]{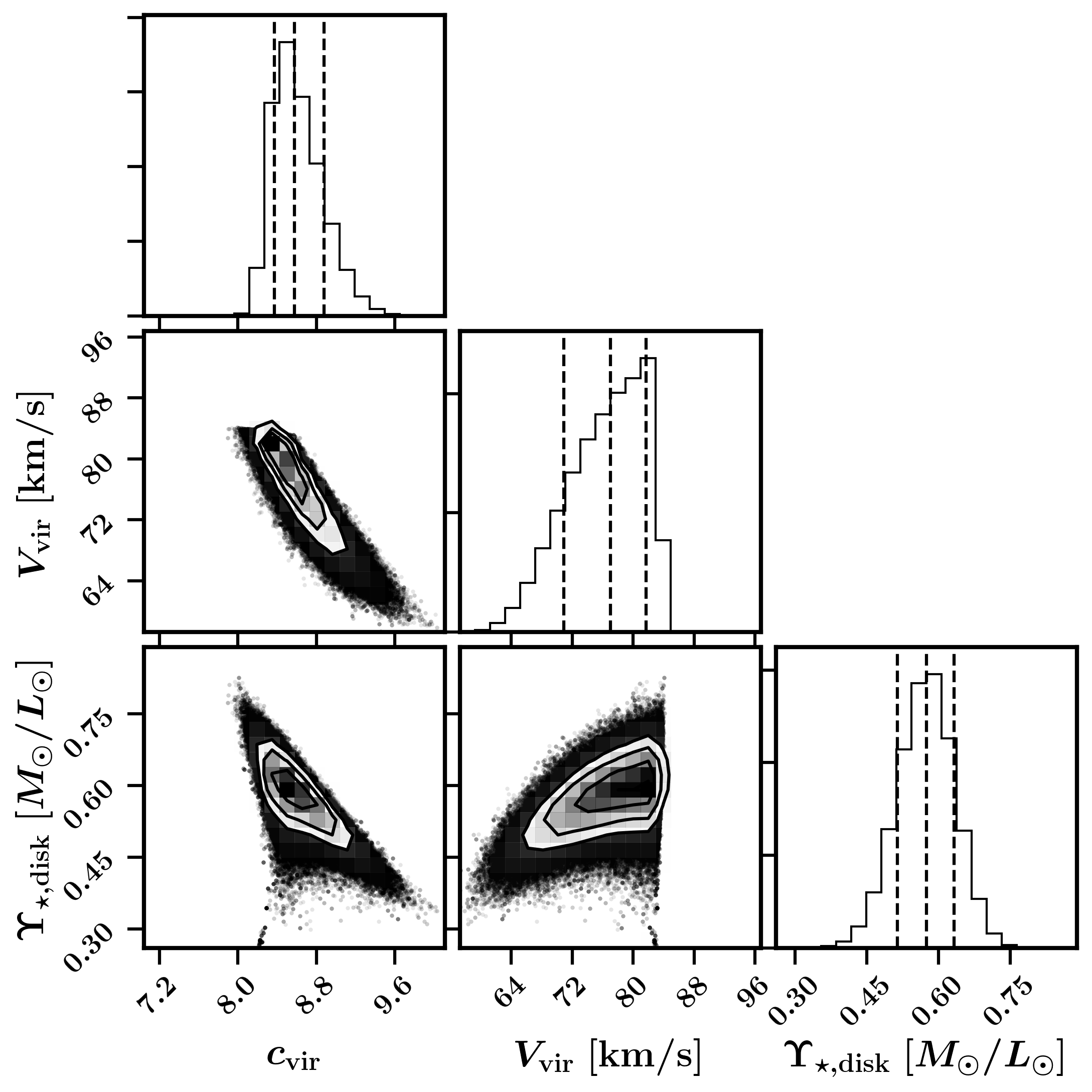}
    \includegraphics[width=0.3\textwidth]{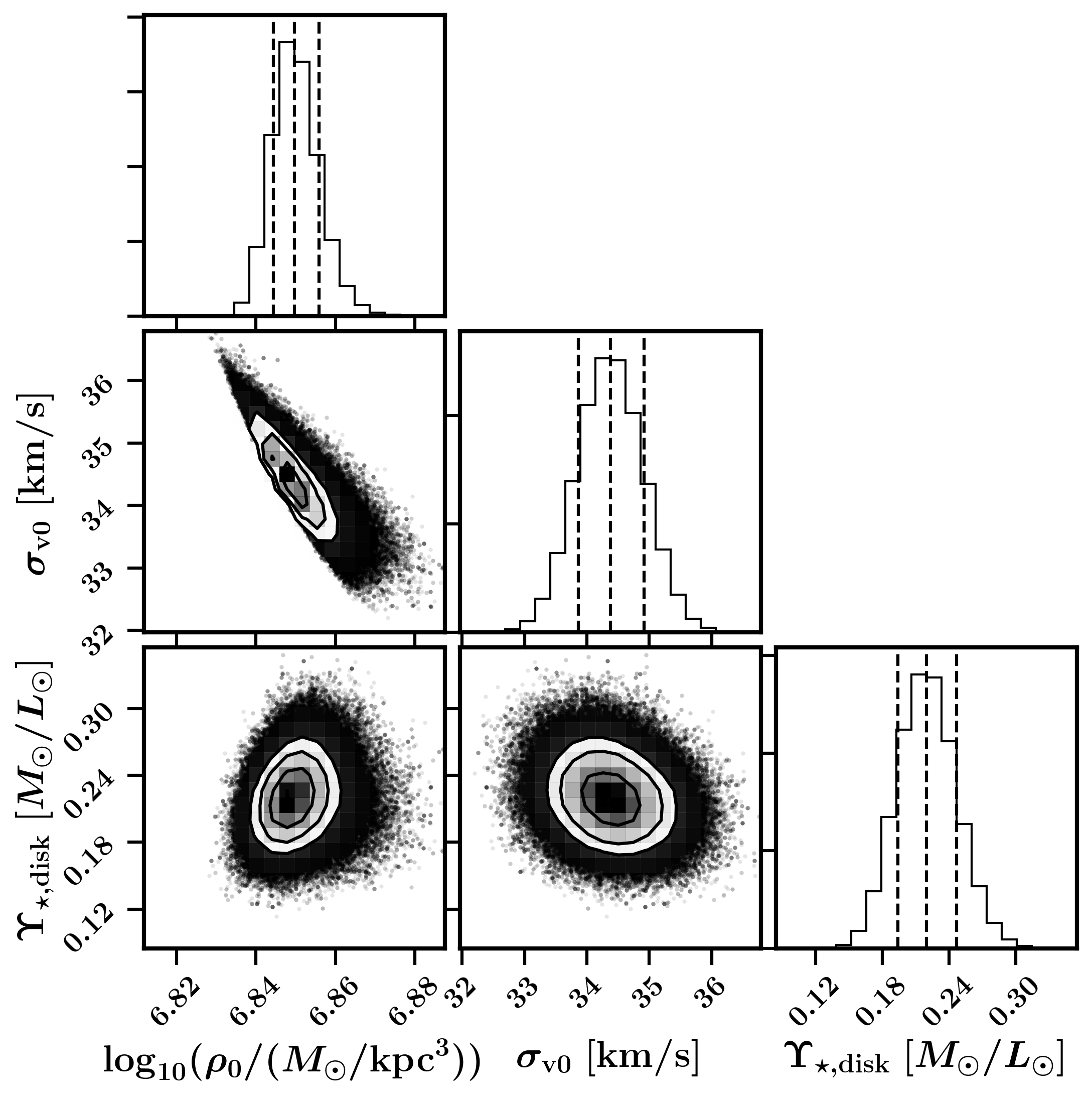}
    \caption{NFW~(left column), DC14~(middle column), and SIDM~(right column) fits to the IC2574 rotation curve with corresponding corner plots.}
\end{figure}

\begin{figure}[h!]
    \centering
    \includegraphics[width=0.3\textwidth]{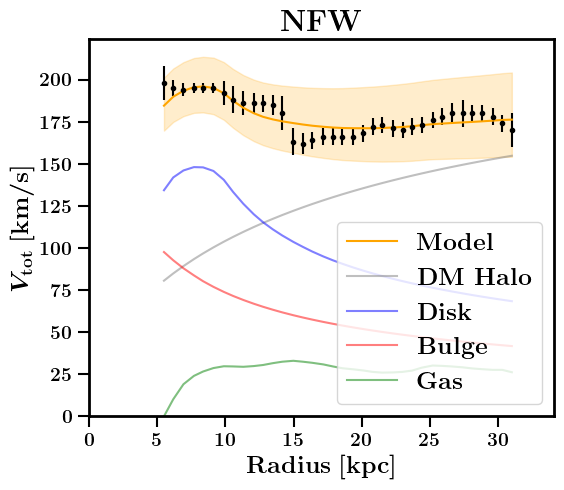}
    \includegraphics[width=0.3\textwidth]{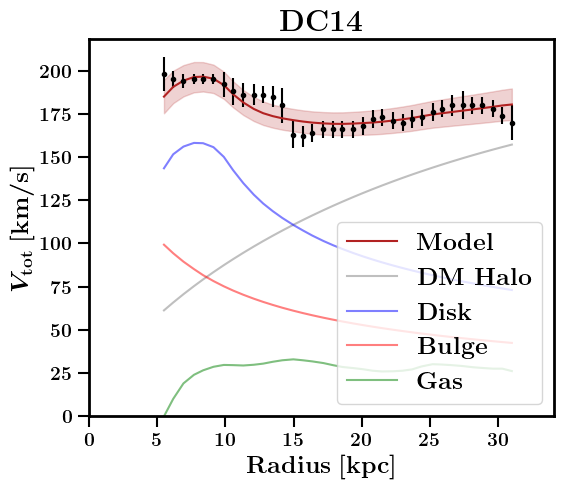}
    \includegraphics[width=0.3\textwidth]{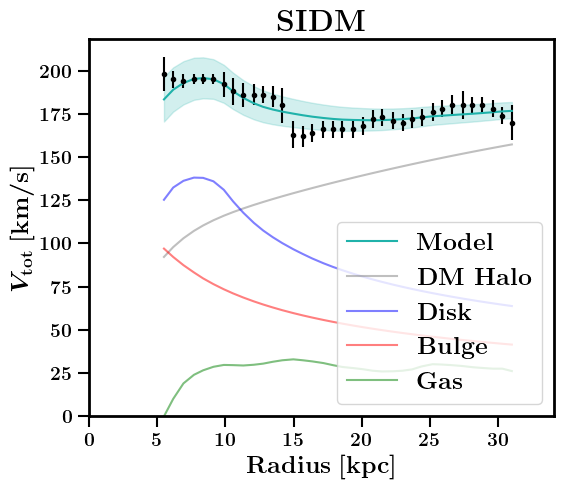}
    \includegraphics[width=0.3\textwidth]{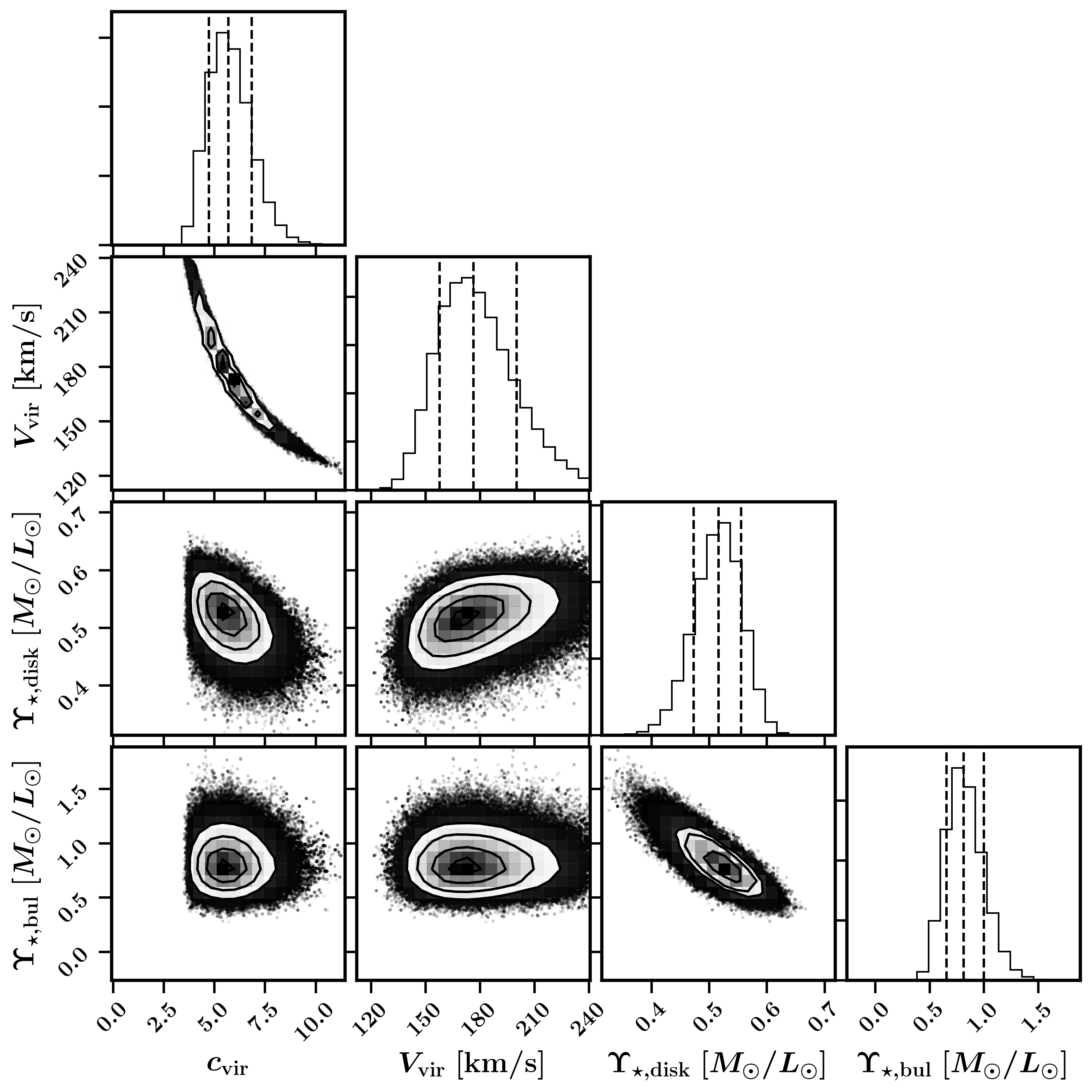}
    \includegraphics[width=0.3\textwidth]{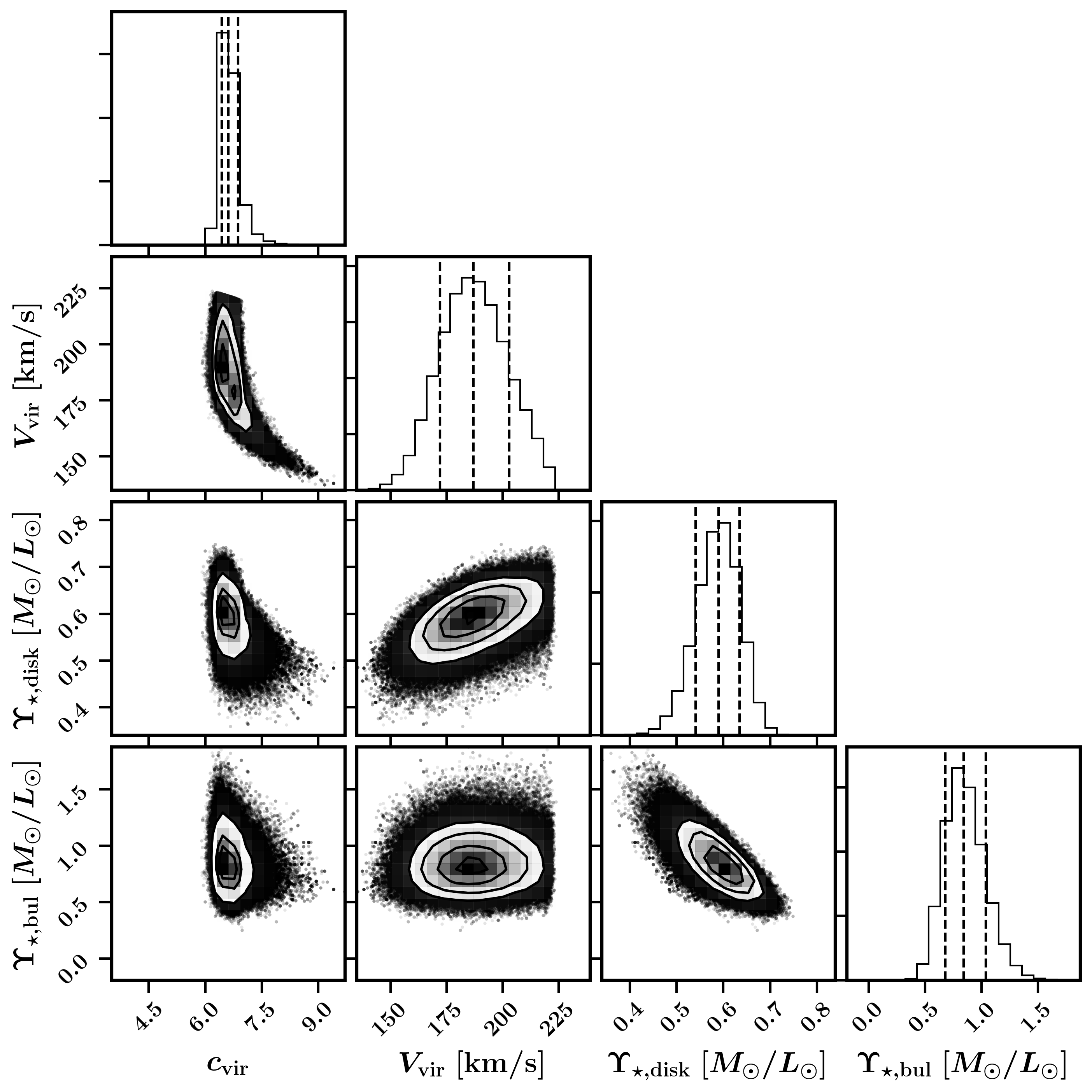}
    \includegraphics[width=0.3\textwidth]{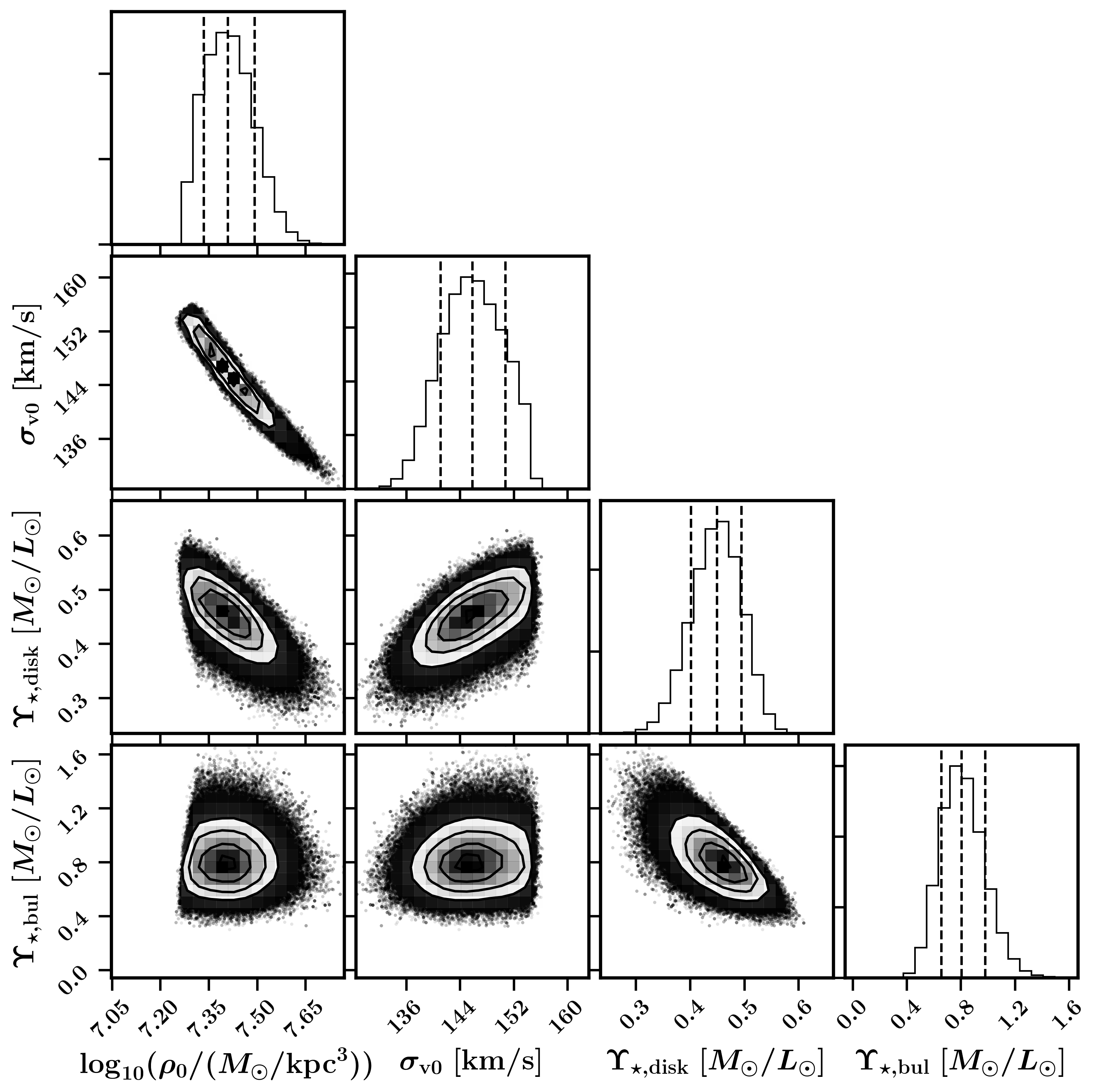}
    \caption{NFW~(left column), DC14~(middle column), and SIDM~(right column) fits to the NGC4013 rotation curve with corresponding corner plots.}
\end{figure}

\clearpage

\section{Outlier Galaxies}
\label{app:B}

\begin{figure}[h!]
    \centering
    \includegraphics[width=0.9\textwidth]{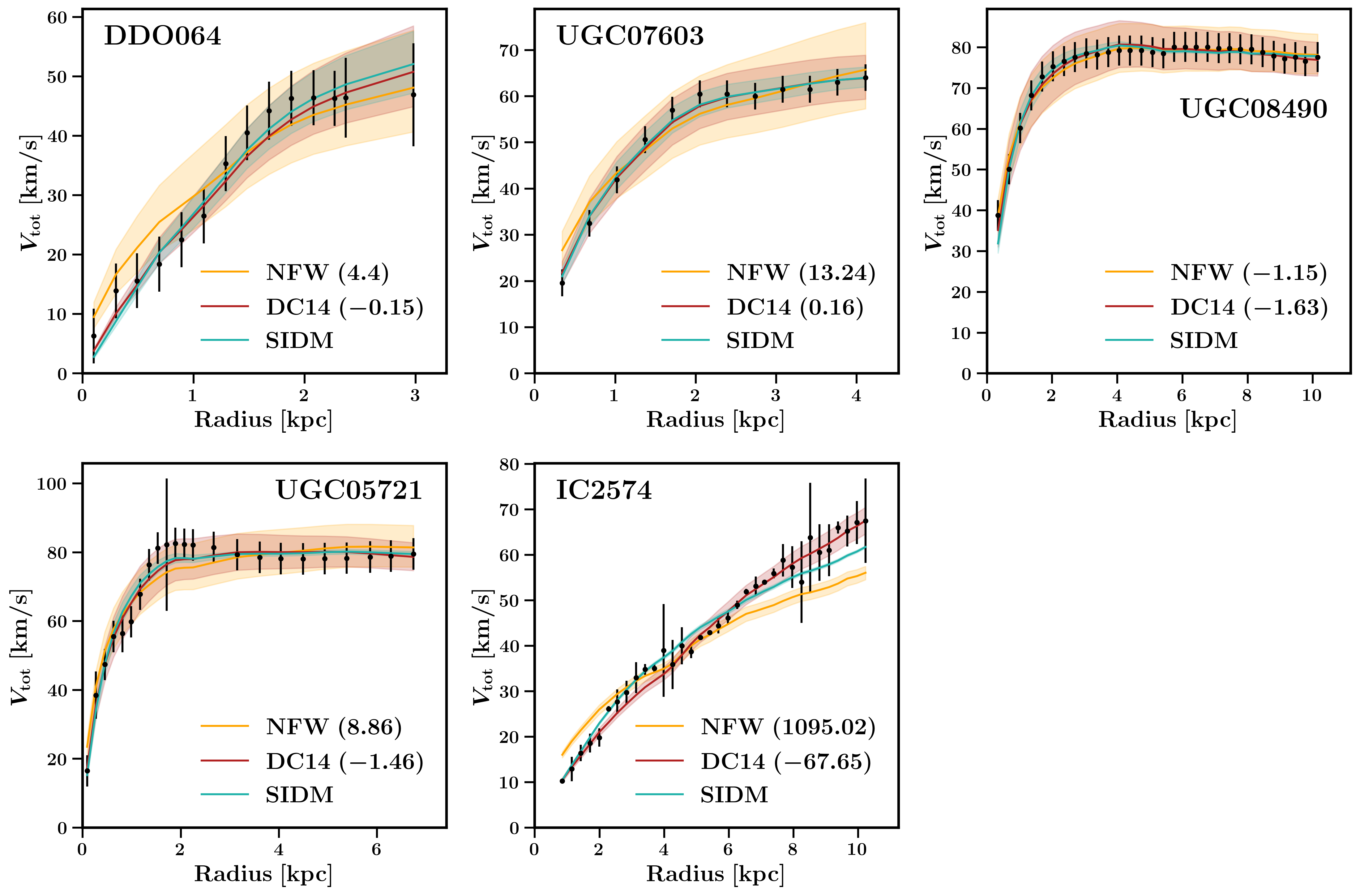}
    \caption{Best-fit rotation curves for galaxies which are both in our sample and are marked as outliers in~\cite{2018MNRAS.473.4392S}. The insets within each of the panels provide values for $\Delta$BIC between NFW and SIDM, and between DC14 and SIDM. Positive values correspond to a preference for SIDM over the CDM model.}
    \label{fig:Santos_RCs}
\end{figure}

\begin{figure}[h!]
    \centering
    \includegraphics[width=0.9\textwidth]{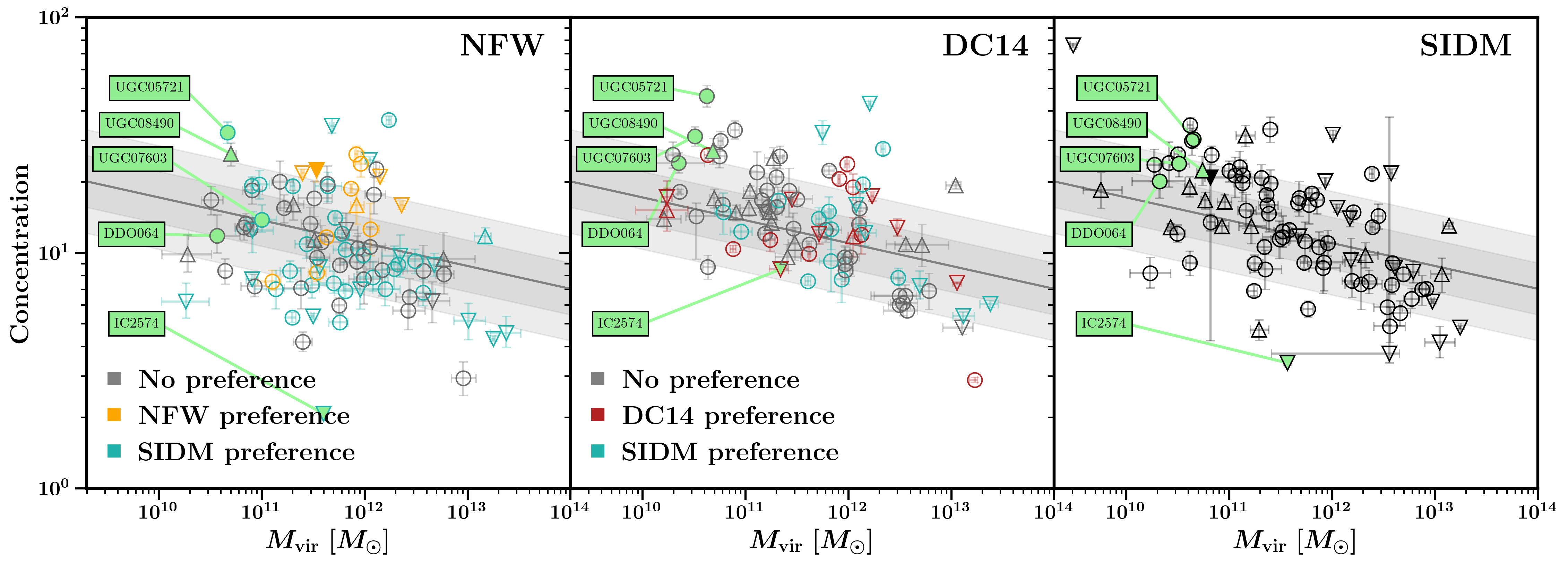}
    \caption{Concentration relations for galaxies in our sample with those marked as outliers by~\cite{2018MNRAS.473.4392S} shown in green.}
    \label{fig:Santos_Concentrations}
\end{figure}

This appendix provides the best-fit rotation curves (Fig.~\ref{fig:Santos_RCs}) and concentration values (Fig.~\ref{fig:Santos_Concentrations}) for galaxies that are both in our sample and are marked as outliers in~\cite{2018MNRAS.473.4392S}. Specifically, these are galaxies which seem to have values of the ratio of $V_{\rm tot}$ at 2 kpc to its value at the last measured radius that are not reproduced in NIHAO simulations. The inset within each rotation curve provide values for $\Delta$BIC between NFW and SIDM, and between DC14 and SIDM. We find that only one out of the 5 galaxies (IC2574) shows a strong preference for DC14 over SIDM, with the other showing no strong preference between these two cases. The other galaxies have $\Delta$BIC values too small to make any conclusive claims. Additionally, as shown in Fig.~\ref{fig:Santos_Concentrations}, we find that these galaxies tend to be outliers in terms of their concentration relations. This could potentially explain why simulations find it challenging to reproduce galaxies of this type.

\clearpage
\bibliographystyle{aasjournal}
\bibliography{mainCitations}

\end{document}